\pdfoutput=1
\documentclass[fleqn,usenatbib]{mnras}

\usepackage{newtxtext,newtxmath}
\usepackage[T1]{fontenc}

\DeclareRobustCommand{\VAN}[3]{#2}
\let\VANthebibliography\thebibliography
\def\thebibliography{\DeclareRobustCommand{\VAN}[3]{##3}\VANthebibliography}

\usepackage{graphicx}	% Including figure files
\usepackage{amsmath}	% Advanced maths commands
\title[The MFP approximation and stellar collisions]{The mean free path approximation and stellar collisions
in star clusters: Numerical exploration of the analytic rates and the role of perturbations on binary star mergers}

\author[B. Reinoso et al.]{Basti\'an Reinoso$^{1}$\thanks{E-mail: bastian.reinoso@uni-heidelberg.de},
Nathan W. C. Leigh$^{2,3}$, Carlos M. Barrera-Retamal$^{2}$, 
\newauthor Dominik Schleicher$^{2}$, Ralf S. Klessen$^{1,4}$, and Amelia M. Stutz$^{2,5}$
\\
$^{1}$Universit\"at Heidelberg, Zentrum f\"ur Astronomie, Institut f\"ur Theoretische Astrophysik, Albert-Ueberle-Str. 2, 69120 Heidelberg, Germany \\
$^{2}$Departamento de Astronom\'ia, Facultad Ciencias F\'isicas y Matem\'aticas, Universidad de Concepci\'on, Av. Esteban Iturra s/n Barrio Universitario,\\ Casilla 160-C, Concepci\'on, Chile \\
$^{3}$Department of Astrophysics, American Museum of Natural History, New York, NY 10024, USA \\
$^{4}$Universit\"{a}t Heidelberg, Interdisziplin\"{a}res Zentrum f\"{u}r Wissenschaftliches Rechnen, Im Neuenheimer Feld 205, 69120 Heidelberg, Germany \\
$^{5}$Max-Planck-Institute for Astronomy, Konigstuhl 17, 69117 Heidelberg, Germany
}

\date{Accepted XXX. Received YYY; in original form ZZZ}

\pubyear{2015}

\begin{document}
\label{firstpage}
\pagerange{\pageref{firstpage}--\pageref{lastpage}}
\maketitle

\begin{abstract}
In this paper we compute predictions for the number of stellar collisions derived from analytic models based on the mean free path (MFP) approximation and compare them to the results of $N$-body simulations.  Our goal is to identify the cluster conditions under which the MFP approximation remains valid.  Adopting a range of particle numbers ($100\leq N\leq5000$) and different combinations of particle masses and radii, we explore three different channels leading to stellar collisions, all of which are expected to occur in realistic stellar environments.
At high densities, binaries form from isolated three-body interactions of single stars. Hence, we consider collisions between single stars and collisions involving binary stars, after they form in our simulations.  For the latter, we consider two channels for mergers, namely direct stellar collisions during chaotic single-binary interactions and perturbation-driven mergers of binaries due to random walks in eccentricity approaching unity. In the densest systems considered here, a very massive object is formed at the cluster centre, causing local stellar orbits to become increasingly Keplerian and the assumptions going into our analytic model to break down. Before reaching this limit, we obtain excellent agreement between our theoretical predictions and the simulations:  the analytic rates are typically accurate to within one standard deviation for the entire parameter space considered here, but the agreement is best for short integration times. Our results have direct implications for blue straggler formation in dense star clusters, and stellar mergers in galactic nuclei hosting massive black holes.
\end{abstract}

%and as this object grows this should mark the transition from a chaotic to a more deterministic dynamical evolution, as its contribution to the total gravitational potential increases and the local stellar orbits become increasingly Keplerian. In the limit of large central object masses, the influence radius becomes large and the stellar orbits are typically assumed not to be changing significantly over short time scales, as occurs for low-number chaotic systems.  For our simulations, however, we do not reach this limit, suggesting that perturbative encounters remain important for a large fraction of the relevant parameter space. 


%%%%%%%%%%%%%%%%%%%%%%%%%%%%%%%%%%%%%%%%%%%%%%%%%%

%%%%%%%%%%%%%%%%% BODY OF PAPER %%%%%%%%%%%%%%%%%%

\section{Introduction}

The mean free path (MFP) approximation has been widely used throughout the field of astrophysical dynamics for centuries. Simple estimates based on the stellar number density, the collisional cross section, and the stellar velocities (``n$\sigma$v'') for the rate of direct collisions between pairs of particles can be, and very often are, applied to a wide variety of astrophysical problems \citep[e.g.][]{binney87}, ranging from direct stellar collisions in star clusters and galactic nuclei \citep[e.g.][]{fregeau04,Portegies2004,Naoz18}, the growth of planetary embryos in protoplanetary disks \citep[e.g.][]{goldreich82}, tidal stripping during interactions between pairs of galaxies \citep[e.g.][]{ogiya18}, and even atoms and molecules colliding in gas clouds \citep[e.g.][]{spitzer41a,spitzer41b,spitzer42}.  Typically, the ``sticky-star approximation'' is adopted to compute the relevant rates and timescales, where a collision is defined as occurring when the radii of two or more stars overlap in both time and space.

%PARAGRAPH ABOUT THE LACK OF SIMULATIONS FOR THIS, MOSTLY DONE FOR STARS IN CLUSTERS - REF ME AND FREGEAU.
%FIND GONJI'S PAPER, and the gaburov papers!!!  ACTUALLY, INVITE SIMON TO COLLABORATE, and help me fill in refs since he did a lot of the work in thie field!!!
To date, few studies have considered the accuracy of analytic methods for computing the rates of particle collisions in realistic astrophysical environments \citep[e.g.][]{hut83a,hut83b,fregeau04}.  Most of the work that has been done focused on direct collisions between pairs of objects (galaxies, stars, planets, etc.) during isolated small-number chaotic interactions mediated by gravity.  For example, \citet{leigh12} studied the probability of collisions occurring during gravitationally-bound small-number ($N =$ 3, 4, 5, 6) chaotic interactions involving identical finite-sized particles.  The authors showed that the collision probability scales approximately as $N^2$, as expected from combinatorics and the MFP approximation (i.e. the collision probability should be proportional to {$N \choose 2$}).  In subsequent studies \citet{leigh15} and  \citet{,leigh17,leigh18} expanded the parameter space to consider particles having different masses and radii.  Eventually, they were able to build from first principles, on a combinatorics-based back bone, analytic predictions for the relative probabilities of different collision scenarios occurring (e.g., multi-collision scenarios).  Using numerical simulations, they confirmed the validity of their method and demonstrated its robustness for any number of interacting particles with any combinations of particle masses and radii.  

Other studies using numerical simulations of collisions involving stars, typically focused on cross-sections, since these can be inserted directly into the standard ``n$\sigma$v'' timescales derived using the MFP approximation \citep[e.g.][]{hut83a,hut83b,fregeau04,fregeau06}.  But these too only considered small numbers of particles, typically four or less.
In a recent study \cite{Barrera20} compared analytic collision timescales to collision times obtained from $N$-body simulations.  They considered a range of particle numbers ($N\sim10^3$) and included particles with different masses and radii. They showed that the collision times typically agree to within one standard deviation with the simulated results, and that the agreement is better for smaller $N$ and a narrow mass spectrum.

In astrophysics, runaway collisions have been first suggested to be relevant in dense young star clusters and globular clusters \citep{Portegies2002, Portegies2004}. \citet{Baumgardt2011} explored them particularly in the context of young clusters, as a potential origin of very massive stars. In the early Universe, runaway collisions in dense stellar clusters have been considered as a potential origin of massive black hole seeds, employing analytic relations \citep{Devecchi2009, Devecchi2012} as well as numerical simulations \citep{Katz2015, Sakurai2017,Sakurai2019, Reinoso2018, Reinoso20}. Collisions in clusters of stellar-mass black holes were suggested to be relevant for the formation of massive black holes by \citet{Lupi2014}. Furthermore, the interplay between collisions and accretion appears as a particularly promising mechanism for the formation of the first supermassive black hole seeds \citep{Boekholt2018, Alister2020,Tagawa20,Chon20,Das2020}.
Stellar collisions are also invoked to explain the formation of exotic stellar populations, such as blue stragglers and the S-stars, a group of high mass stars distributed in a disk-like structure very close ($\lesssim 0.04$~pc) to Sgr~A$^{*}$ in the Galactic centre \citep[e.g.][]{Eckart1997,Ghez2003,Naoz18}.
Both blue stragglers and the S-stars are thought to be products of stellar mergers between main-sequence stars.
%(see Section~\ref{Sstars} for more details).}

It is important to understand how good is the ``$n\sigma v$'' approximation in extreme environments where frequent stellar collisions are expected to occur. The approximation may not be valid because cluster conditions are not constant over time. Two-body relaxation causes the core radius to shrink and the central stellar density to increase, and the core velocity dispersion along with it. Slower heavier stars that collide outside the core begin to drift into it via two-body relaxation, populating the cluster centre and subsequently transferring kinetic energy to other stars, having a direct impact on the velocity distribution of stars in both the core and the halo. If collisions are frequent, the mean stellar mass will also change, and more massive collision products have larger cross sections for collisions. Furthermore the hard-soft boundary varies as the masses of the binary components change and the core contracts to become hotter with a higher velocity dispersion, which would modify the timescale for single-binary interactions. It may not be a good approximation to take then the average semi-major axis when computing the single-binary collision timescale, but the only way to know for sure is to perform the numerical experiments and compare to the analytic theory, as we do in this paper. The presence of collision products will also change the properties of binary systems since during interactions with single stars, they tend to retain higher mass stars and eject the lightest one.  This will cause rapid evolution in the stellar mass function at both the high-mass and low-mass ends. 
%Quantifying these competing timescales and the rate of evolution of the stellar and binary mass functions is one of the goals of this paper.}
%These are some of the reasons why the ``$n\sigma v$'' approximation may fail to estimate the number of collisions in a star cluster. 

In this paper we present a comparison between the number of stellar collisions obtained from $N$-body simulations of dense star clusters and the predicted number of collisions from analytic collision rates. We show that the predicted and simulated values agree to within a factor of order unity. We identify in our simulations collisions of binary stars that occur due to perturbations coming from single stars that pass close to the binary on hyperbolic orbits. We discuss the implications of this process for the formation of stellar exotica in realistic stellar environments.

The manuscript is organized as follows. We describe our $N$-body and analytic models in Sec.~\ref{sec:methods}, and present the results of their comparisons in Sec.~\ref{sec:results}. We discuss the applicability of our results to the formation of stellar exotica in Sec.~\ref{sec:applications}. A final discussion and summary is presented in Sec.~\ref{sec:discussion}.

\section{Methods} \label{sec:methods}

We present in this section the $N$-body models that served as initial conditions for the numerical simulations used in this paper. We also discuss how we count the number of collisions from our simulated data. Then we present the analytic model and describe how we compute the predicted number of stellar collisions. We focus our analysis to the cluster core and thus all the collisions presented in this paper (both from simulations and analytic rates) refer to collisions occuring in the core.
%We further , for direct comparison to the analytic predictions in Section~\ref{sec:results}.  

\subsection{Initial conditions}

In our study we model compact star clusters assuming a Plummer distribution \citep{Plummer1911} with virial radius $R_{\rm v}$ = 0.14~pc.  Every system is composed of identical stars initially having the same masses and radii, with a total stellar mass of $M_{\rm cluster} = 10^4$~M$_{\odot}$. With these properties fixed, we vary the total number of stars $N$ and the stellar radius $R_{\rm star}$ to produce 24 different $N$-body models, all of which are listed in Table~\ref{tab:IC_sims}. The initial masses of the stars are decided as $m_{\rm ini}=M_{\rm cluster}/N$. For each of the 24 models mentioned above, we run 6 simulations varying the initial random seed, yielding a total of 144 $N$-body simulations.

Every cluster begins in a state of virial equilibrium, and is left to evolve for 2000 $N$-body time units which equals 15.6~Myr in the lifetimes of our simulated clusters.  We are here interested only in the first stages of the cluster evolution, as explained more clearly below (see fourth column in Table~\ref{tab:list_of_sims}).

\begin{table}
	\centering
	\caption{Initial conditions for each $N$-body model. We perform 6 simulations per model.}
	\label{tab:IC_sims}
	\begin{tabular}{l r r r r c c} % four columns, alignment for each
		\hline
		Model & $N$ & $m_{\rm ini}$ & $R_{\rm star}$ & $r_{\rm core}$ & $v_{\rm rms}$  & $n_{\rm core}$  \\
		%\hline
		 &  & (M$_\odot$) & (R$_\odot$) & (pc) & (km s$^{-1}$) & ($10^6$ pc$^{-3}$) \\
		\hline
1 & 100 & 100 & 20 & 0.042 & 17.450 & 0.318 \\ 
2 & 100 & 100 & 50 & 0.042 & 16.865 & 0.327 \\ 
3 & 100 & 100 & 100 & 0.045 & 11.818 & 0.266 \\ 
4 & 100 & 100 & 200 & 0.058 & 12.063 & 0.124 \\ 
5 & 100 & 100 & 500 & 0.061 & 12.194 & 0.108 \\ 
6 & 100 & 100 & 1000 & 0.054 & 13.017 & 0.152 \\
\\
7 & 500 & 20 & 20 & 0.040 & 16.618 & 1.930 \\ 
8 & 500 & 20 & 50 & 0.042 & 15.804 & 1.579 \\ 
9 & 500 & 20 & 100 & 0.045 & 15.501 & 1.320 \\ 
10 & 500 & 20 & 200 & 0.050 & 15.673 & 0.967 \\ 
11 & 500 & 20 & 500 & 0.047 & 15.553 & 1.126 \\ 
12 & 500 & 20 & 1000 & 0.040 & 15.678 & 1.864 \\
\\
13 & 1000 & 10 & 20 & 0.039 & 15.037 & 3.886 \\ 
14 & 1000 & 10 & 50 & 0.047 & 15.640 & 2.252 \\ 
15 & 1000 & 10 & 100 & 0.045 & 15.926 & 2.550 \\ 
16 & 1000 & 10 & 200 & 0.044 & 15.648 & 2.751 \\ 
17 & 1000 & 10 & 500 & 0.048 & 15.099 & 2.190 \\ 
18 & 1000 & 10 & 1000 & 0.040 & 14.941 & 3.780 \\
\\
19 & 5000 & 2 & 20 & 0.042 & 15.099 & 16.014 \\ 
20 & 5000 & 2 & 50 & 0.042 & 15.120 & 16.014 \\ 
21 & 5000 & 2 & 100 & 0.043 & 15.095 & 15.153 \\ 
22 & 5000 & 2 & 200 & 0.043 & 15.116 & 15.258 \\ 
23 & 5000 & 2 & 500 & 0.043 & 15.117 & 15.153 \\ 
24 & 5000 & 2 & 1000 & 0.043 & 15.097 & 15.258 \\ 
		\hline
	\end{tabular}
\end{table}

\subsection{Numerical simulations}
\label{sec:numerical_simulations}

All the simulations presented in this paper were performed with {\small NBODY6} \citep{Aarseth2000}, a direct $N$-body integrator which makes use of the fourth order Hermite method, block time-steps, KS regularization for treating close encounters \citep{Kustaanheimo1965,Mikkola1998}, and a spatial hierarchy for the force computation \citep{AhmadCohen73}.
Stellar collisions are detected once the radii of particles overlap in both space and time during the integration, and they are replaced by a new single particle. This new particle is placed at the centre of mass of the previous configuration. The mass and velocity are computed assuming mass and linear momentum conservation. The new radius is calculated by assuming that the density of the progenitors and the collision product are the same. 
For simplicity, we do not include in our simulations stellar evolution nor tidal interactions between stars.

\subsection{Counting the number of collisions from the simulations}
\label{sec:col_from_sims}

The majority of the collisions in our simulations happen in the cluster core (see below), or very close to it, therefore we focus our analysis on the central regions of our clusters. We extract the required information about stellar collisions from the output files and snapshots of {\small NBODY6} and select only those collisions occurring inside the 10\% Lagrangian radius, which we will henceforth refer to as the core radius $r_{\rm core}$.
The snapshot output frequency is 1 $N$-body unit of time, which corresponds to 7800~yr.

We distinguish between two types of collisions here, according to the classification given by {\small NBODY6}. First, we have both hyperbolic collisions or, equivalently, collisions between stars that are not initially gravitationally bound. Second, we have binary-mediated collisions or collisions between two stars that become gravitationally bound before the collision event, and these mergers/collisions can be decomposed into two groups - mergers of the binary system mediated by perturbations from bound single stars, and collisions mediated by perturbations from unbound single stars (sometimes undergoing prolonged chaotic interactions, with the eccentricity doing a random walk to higher and higher values, before a merger occurs when the stellar radii overlap at pericentre).

We present for each $N$-body model, the average number of hyperbolic ${N}_{\rm Hyp}$ and binary-mediated ${N}_{\rm Bin}$ collisions in Table~\ref{tab:list_of_sims}, as well as the average total number of collisions ${N}_{\rm col,sim}$. The errors are computed assuming Poisson statistics.

\subsection{The analytic model}
The analytic model that we use in this paper is constructed from the mean times or rates between stellar encounters derived from the MFP approximation. In a similar way to \cite{Leigh2011}, we include single-single and single-binary interactions via encounter timescales, but we include the gravitationally focused cross sections in the derivation (see Appendix~\ref{sec:timescales_gravfoc}), which are applicable to the cores of star clusters. Consequently the mean time between single-single encounters in the core is given by:
 \begin{eqnarray}
 \label{eq:tau_1+1}
\tau_{1+1} &=& 8.3\times 10^{13} \ (1-f_b-f_t)^{-2} \left( \frac{10^3\ \rm pc^{-3}}{2n_{\rm core}}\right)^2   \times \\ 
 & &\left( \frac{1 \ \rm pc}{r_{\rm core}}\right)^3  \left( \frac{5\ \rm km\ s^{-1}}{v_{\rm rms}}\right) \left( \frac{0.5 \ \rm R_\odot}{\langle R \rangle}\right)^2 \times \nonumber \\
 & & \left[ 1+ 7635\left( \frac{\langle m \rangle}{0.5 \rm \ M_\odot}\right) \left( \frac{0.5 \rm \ R_\odot}{\langle R \rangle}\right) \left( \frac{5\ \rm km\ s^{-1}}{v_{\rm rms}}\right)^2 \right]^{-1} \rm yr, \nonumber
\end{eqnarray}
%\begin{eqnarray}
%\label{eq:tau_1+1}
%    \tau_{1+1} &=& 1.1\times 10^{10} (1-f_b)^{-2} \left( \frac{1~ {\rm pc}}{r_{\rm core}} \right)^3 \left( \frac{10^3 ~{\rm pc}^{-3}}{n_{\rm core}} \right)^2  \nonumber \\
%    & & \left( \frac{v_{\rm rms}}{5 ~{\rm km~s^{-1}}} \right) \left( \frac{0.5~{\rm M_\odot}}{\langle m\rangle} \right) \left( \frac{0.5~{\rm R_\odot}}{\langle R \rangle} \right)~{\rm yr},
%\end{eqnarray}
where $f_b$ is the fraction of binary systems in the core, defined as $f_b = N_b/N_{\rm core}$, where $N_{b}$ is the number of binaries in the core and $N_{\rm core}$ is the total number of objects in the core (i.e., including both singles and binaries), $r_{\rm core}$ and $n_{\rm core}$ are the core radius and core number density, $v_{\rm rms}$ is the root-mean-square velocity,
$\langle m \rangle $ and $\langle R \rangle$ are the mean stellar mass and mean stellar radius.

The mean time between single-binary interactions is given by:
\begin{eqnarray}
\label{eq:tau_1+2}
\tau_{1+2} &=& 1.8\times 10^9 \ (1-f_b-f_t)^{-1} \ f_b^{-1} \left( \frac{10^3\ \rm pc^{-3}}{2n_{\rm core}}\right)^2  \times \\
& & \left( \frac{1 \ \rm pc}{r_{\rm core}}\right)^3 \left( \frac{5\ \rm km\ s^{-1}}{v_{\rm rms}}\right) \left( \frac{1 \ \rm AU}{a_b}\right)^2 \times \nonumber \\
 & & \left[ 1+ 53\left( \frac{\langle m \rangle }{0.5 \rm \ M_\odot}\right) \left( \frac{1 \rm \ AU}{a_b}\right) \left( \frac{5\ \rm km\ s^{-1}}{v_{\rm rms}}\right)^2 \right]^{-1} \rm yr, \nonumber
\end{eqnarray}
%\begin{eqnarray}
%\label{eq:tau_1+2}
%    \tau_{1+2} &=& 3.4\times 10^{7} \left(1-f_b\right)^{-1} f_b^{-1} \left( \frac{1~{\rm pc}}{r_{\rm core}} \right)^3 \left( \frac{10^3 ~{\rm pc^{-3}}}{n_{\rm core}} \right)^2  \nonumber \\
%    & & \left( \frac{v_{\rm rms}}{5 ~{\rm km~s^{-1}}} \right) \left( \frac{0.5~{\rm M_\odot}}{{\langle m \rangle}} \right) \left( \frac{1~{\rm AU}}{a_b} \right)~{\rm yr},
%\end{eqnarray}
where $a_b$ is the mean semi major axis for binaries.

We note that binary-binary collisions can be ignored, due to the very low binary fractions in our simulations and the fact that single-binary interactions dominate over binary-binary interactions for $f_{b}$~$\lesssim$~0.1 \citep{Leigh2011}.

We also attempted a different model in which we use Eq.(\ref{eq:1+1}) to estimate the number of single-single collisions between equal mass stars, combined with the gravitationally focused encounter rate for unequal mass stars presented in \cite{leigh17}. A similar experiment was done in \cite{Barrera20}. This model however overestimates the number of collisions between unequal mass stars by a factor of $\sim2$.

\subsection{Calculating the predicted number of collisions from the analytic model}
\begin{figure}
	\includegraphics[width=\columnwidth]{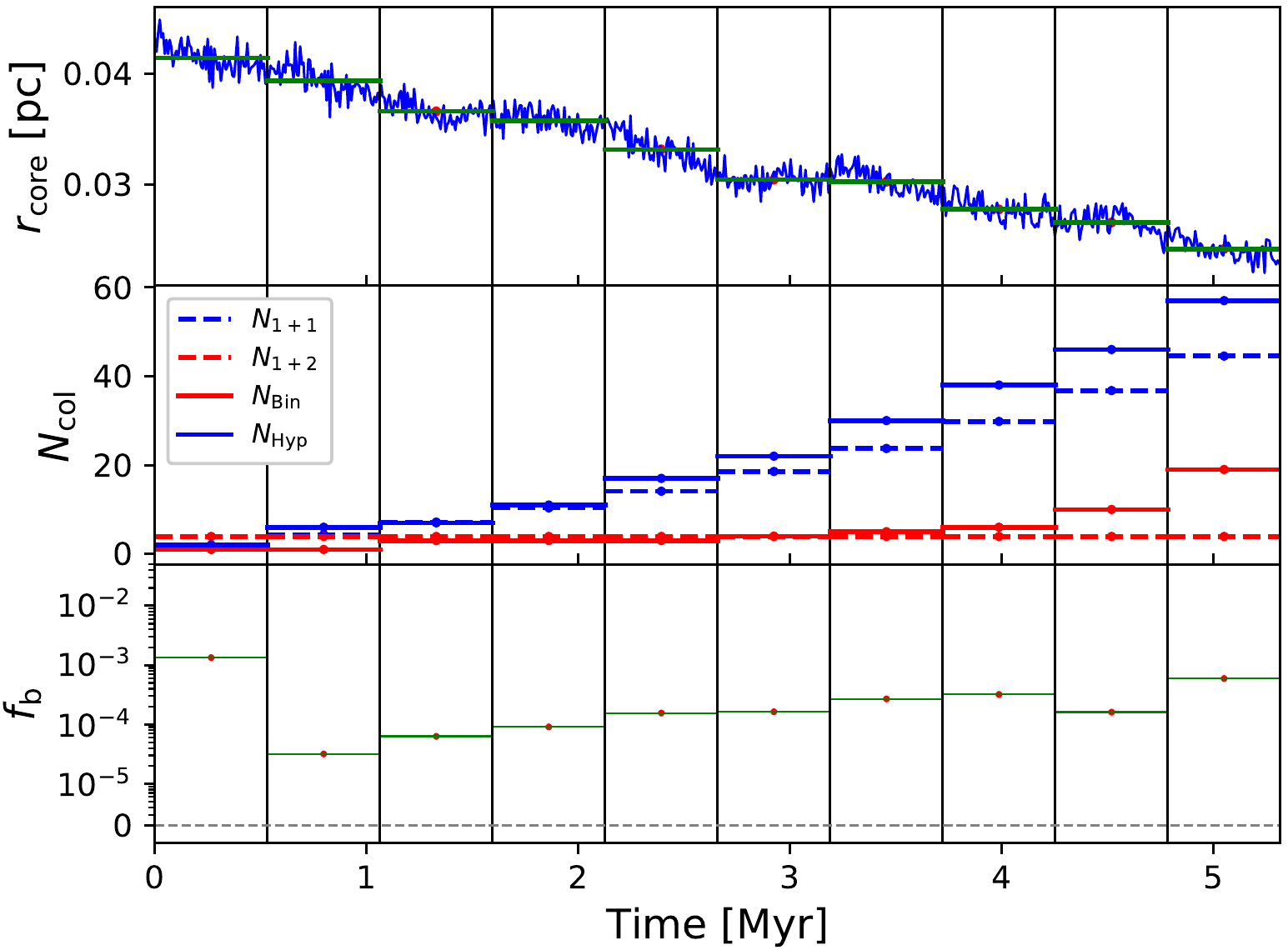}
    \caption{Illustration of the calculation for the number of collisions from our analytic model. We present, in the top panel, the evolution of the core radius. The middle panel shows the number of calculated 1+1 and 1+2 collisions as described in Sec.~\ref{sec:cols_from_analytic_model}. We also show the number of hyperbolic and binary collisions counted from the simulation for comparison. The bottom panel shows the fraction of binary systems in the core.
    The black vertical lines shows the partitioning of the simulation time $t_{\rm sim}$ into 10 smaller intervals, each of which contains, on average, 6.8 snapshots. This is one of the simulations of Model 19.}
    
    %We present, in the top panel, the evolution of the core radius, along with the partitioning of the time lapse $\Delta t$ into 10 smaller intervals with vertical black lines. In each interval we calculate the core radius by averaging over simulation snapshots. We compute the accumulated number of 1+1 (blue dashed) and 1+2 (red dashed) collisions from Eq.~(\ref{eq:1+1}) and (\ref{eq:N_1+2}), and present them in the middle panel. We also plot the accumulated number of hyperbolic (solid blue) and binary collisions (solid red), counted from the simulation.
    %In the bottom panel, we present the fraction of binary systems at each time interval, calculated as the mean number of binary systems, divided by the mean number of stars and averaged over the number of snapshots. There are on average 6.8 snapshots inside each time interval.}
    \label{fig:encounter_rates_example_calculation}
\end{figure}

%and (\ref{eq:AplusB}) 
\label{sec:cols_from_analytic_model}
The quantities going into Eq.~(\ref{eq:tau_1+1}) and (\ref{eq:tau_1+2}) are taken directly from our $N$-body simulations. For this purpose we first choose a simulation time $t_{\rm sim}$ during which the core radius is smoothly decreasing over time, so we avoid going into the stage of core collapse. The simulation times $t_{\rm sim}$ chosen for each model are listed in the fourth column of Table~\ref{tab:list_of_sims}. We partition this time into 10 successive intervals, each with a length $\Delta t=0.1 t_{\rm sim}$ over which we compute the number of collisions expected from our analytic predictions according to averaged cluster core and stellar properties in that interval.
This is illustrated for one of our simulations in Fig.~\ref{fig:encounter_rates_example_calculation} where we show in the top panel the evolution of the core radius and the partitioning of the simulation time $t_{\rm sim}$ into 10 smaller intervals via vertical black lines. The horizontal green lines in each interval mark the average core radius in the top panel and the mean binary fraction in the bottom panel.
The number of intervals is chosen in such a way that we get representative values for the dynamical properties of the stars in the core by averaging over simulation snapshots, but avoiding large intervals over which the changing properties of the cluster become significant. We check that using 20 intervals does not change our results but using 5 or less intervals leads to lower predicted collision rates.

The total number of collisions expected from our analytic rates is calculated by summing over the number of collisions in each interval $i$:
\begin{eqnarray}
  N_{\rm col, analytic} &=& \displaystyle \sum_{i=1}^{10} N_{1+1,i} + N_{1+2,i} \ ,
\end{eqnarray}
where, we define the number of collisions due to single-single interactions, occurring in the interval $i$, as 
\begin{eqnarray}
\label{eq:1+1}
N_{1+1,i}&=&\frac{\Delta t}{\tau_{1+1,i}}. % + N_{A+B,i}.   
\end{eqnarray}

The number of binary mediated collisions, occurring in the interval $i$, is: 

\begin{eqnarray}
\label{eq:N_1+2}
N_{1+2,i} &=& \left(f_{\rm dir} + \frac{f_{\rm pert}}{n_{\rm pert}}  \right) \frac{\Delta t}{\tau_{1+2,i}}.
\end{eqnarray}
We illustrate this calculation procedure in the middle panel of Fig.~\ref{fig:encounter_rates_example_calculation} by showing with dashed blue and dashed red lines the accumulated number of 1+1 and 1+2 collisions respectively, calculated according to Eq.~(\ref{eq:1+1}) and (\ref{eq:N_1+2}). We also show for comparison the number of hyperbolic and binary collisions counted from the simulations with solid blue and solid red lines respectively. The mean fraction of binaries in the core $f_b$ is presented in the bottom panel.

We note in Eq.~(\ref{eq:N_1+2}) the presence of three correction factors. These are included in order to account for 
two different merger channels as described in Sec.~\ref{sec:binary_mediated_collisions} and \ref{sec:t1_2}.

We emphasize that, since we study binary interactions in the cluster centre, these correction factors apply only to the core.  In order to obtain accurate correction factors outside of this region, we would need to run many more simulations to build up the required statistics, further justifying our choice to focus on the collision rates in the core.
The determination of these factors is described in Sec.~\ref{sec:correction_factors}.

For the calculation of the collision timescale given by Eq.~(\ref{eq:tau_1+1}) and (\ref{eq:tau_1+2}), we compute the values for $r_{\rm core}$, $v_{\rm rms}$, $n_{\rm core}$, $\langle m \rangle$, $\langle R \rangle$, and $f_b$ in every snapshot of our simulations. Then we take the average over the number of snapshots in each interval. We illustrate this for the core radius in the top panel of Fig.~\ref{fig:encounter_rates_example_calculation} and for $f_b$ in the bottom panel. 
The semi-major axis $a_{b}$ is taken to be the mean semi-major axis per binary in the core. 
Typical initial values for the above mentioned quantities for each of our $N$-body models, are presented in Table~\ref{tab:IC_sims} (except for $f_b$ and $a_b$).

%For the calculation of the collision rate between stars of type $A$ and stars of type $B$ from Eq.~(\ref{eq:AplusB}), we compute the following quantities for the stars in the core: the numbers of stars of types $A$ and $B$ or, respectively, $N_{A}$ and $N_{B}$, the mean stellar mass $\langle m \rangle$, the total core stellar mass $M_{\rm core}$, and the total kinetic energy $|E|$ contained in the core. These values are also averaged over the number of snapshots in each time interval.

\subsubsection{Calculating the number of collisions due to (bound and unbound) single-binary interactions}
\label{sec:binary_mediated_collisions}
\label{sec:correction_factors}

The total number of binary mediated stellar collisions is calculated with Eq.~(\ref{eq:N_1+2}) in which we introduce three correction factors.   
We determine those factors by studying the evolution of the binary systems in our simulations. We note however that studying binary evolution requires very frequent snapshot output in order to capture the chaotic and rapid perturbations that a binary experiences from the moment of its formation until disruption or merger. In order to obtain representatives values for the correction factors that we introduce, but at the same time avoiding excessively large data outputs and long computational runtimes, we re-simulate three sets of the 24 models presented in Table~\ref{tab:IC_sims}, with a higher output frequency of 0.01~$N$-body units of time or 78~yr.
These new data are used to compute the parameters $f_{\rm dir}$, $f_{\rm pert}$ and $n_{\rm pert}$.

During a single-binary encounter, a stellar collision between two gravitationally bound stars can occur. When the collision occurs during a bound interaction with a third body we call this type of event a direct binary collision. In order to account for this in our model, we introduce the factor $f_{\rm dir}$, calculated as the number of direct binary collisions divided by the total number of distinct binaries formed in the considered time interval. The values of $f_{\rm dir}$ for each model are presented in Table~\ref{tab:correction_factors}.

\subsubsection{Unbound or perturbative single-binary interactions}
\label{sec:t1_2}
A total of 56 binary collisions in our sample are driven to merger via the cumulative effects of many weak perturbative interactions from passing single stars on hyperbolic orbits relative to the binary centre of mass. This occurs via an exchange of orbital energy and angular momentum with the binary, slowly driving it, via a random walk, to smaller orbital separations and higher eccentricities. This process has been studied by means of Monte Carlo methods and simplified encounter rates with field stars \citep{Krolik84,PortegiesZwart1997,Kaib2014}.

In order to model the perturbation-driven mergers, we study the eccentricity and semi-major axis evolution from the time of binary formation until the time of collision/merger. We define a perturbation as a close encounter in which the distance from the centre of mass of the binary to the closest star is less than, or equal to, three times the semi-major axis. Additionally the close encounter should cause an eccentricity change equal to or larger than five per cent.

We introduce the correction factor $f_{\rm pert}$, which accounts for the efficiency of this merger channel, and is calculated as the number of binary systems in which more than 50 per cent of the perturbations are caused by an unbound star, divided by the total number of distinct binaries formed in the considered time interval. For these perturbed binary collisions we define $n_{\rm pert}$ as the number of unbound perturbations. We present the calculated values of $f_{\rm pert}$ and $n_{\rm pert}$ for each model in Table~\ref{tab:correction_factors}.

\begin{figure}
	\includegraphics[width=\columnwidth]{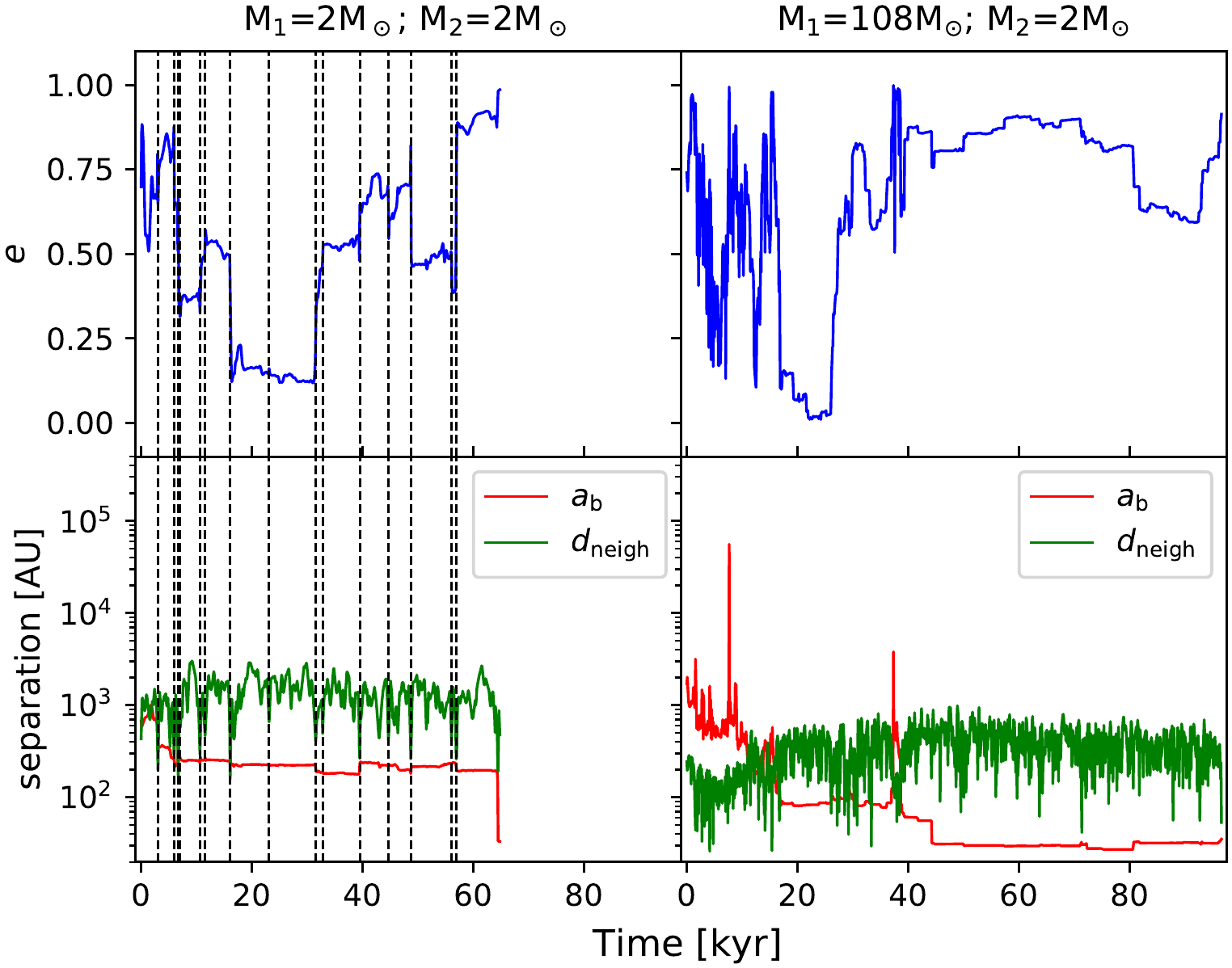}
	\caption{Perturbations to the eccentricity and semi-major axis of two binary systems in our simulations, from the time of binary formation until merger of the binary components. The left panels correspond to a binary with equal mass members whereas the right panels correspond to a binary with the most massive object as a member. The masses are shown at the top of the corresponding panels. We present in the upper panels the time evolution of the eccentricity of the orbit. In the bottom panels, the red line represents the semi-major axis of the binary system and the green line shows the distance separating the binary centre of mass from the closest passing star. The dashed black vertical lines in the left panels mark the perturbations caused by the unbound passing star following the method described in Sec.~\ref{sec:t1_2}. These lines are omitted in the right panels to avoid crowding.}

    \label{fig:binary_perturbation}
\end{figure}

We show in Fig.~\ref{fig:binary_perturbation} two examples of this perturbed binary collision scenario, following the evolution from the time of binary formation until merger. The left panels correspond to a binary with equal mass components perturbed by passing unbound single stars, whereas the right panels correspond to a binary that involves the most massive object in the cluster. In the top panels, we show the time evolution of the eccentricity of the binary orbit.  In the bottom panels, we show the time evolution of the semi-major axis with a red line, and the distance from the centre of mass to the closest third star with a green line. We mark with black vertical lines the times at which a perturbation is detected, as described in the previous paragraph. The mean time between perturbations, in the left panels, is 4074~yr. In this case, perturbations are caused by passing stars that are not bound to the binary. In the right panels perturbations are cause by a third star bound to the binary and are more frequent, with a mean time between perturbations of 367~yr. These perturbations are not shown to avoid crowding of lines.

\section{Results} \label{sec:results}
In this section we confront  the predicted number of stellar collisions derived from our analytic model with the simulated number of collisions obtained from our $N$-body runs.
Our results indicate that the simulated and predicted numbers of collisions agree to within a factor $\lesssim2$ and to within one standard deviation for most of our simulations, as shown in Fig.~\ref{fig:ratio_data_model}. The number of binary-mediated collisions is not well reproduced at later times due to the formation of a massive object at the cluster centre which tends to grow through binary collisions. This is more important in our simulations with $N=5000$, in which a very massive object is always present in the cluster centre at late times. %Our results are consistent with \cite{Barrera20}. 

%\subsection{Collisions occurring pre-binary formation}
\subsection{Hyperbolic collisions}
\begin{figure}
	\includegraphics[width=\columnwidth]{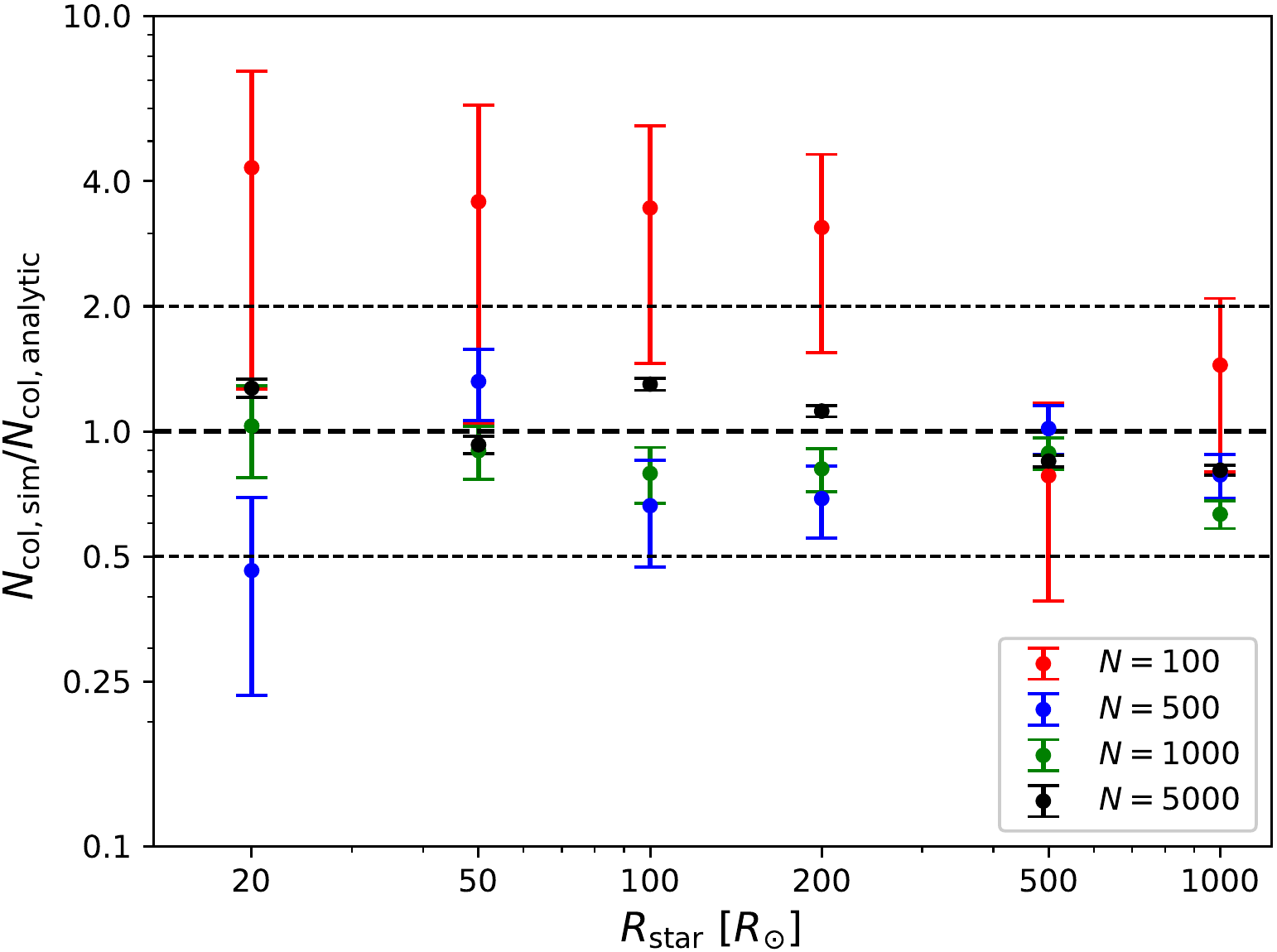}
    \caption{Ratio of simulated to predicted number of collisions as a function of stellar radius. The predicted number of collisions are obtained from our analytic model described in Sec.~\ref{sec:col_from_sims}. The simulated number of collisions are obtained from our $N$-body simulations as described in Sec.~\ref{sec:cols_from_analytic_model}. The data plotted here corresponds to the seventh column in Table  \ref{tab:list_of_sims}.}
    \label{fig:ratio_data_model}
\end{figure}

%In Table~\ref{tab:list_of_sims} we present the results for all our $N$-body models and 
In Fig.~\ref{fig:encounter_rates_example_calculation} we illustrate the calculation procedure for one example simulation. We compare the expected number of collisions over a specified time interval. The middle inset shows the number of simulated hyperbolic, and analytic 1+1 collisions calculated with Eq.~(\ref{eq:1+1}), over each time interval. We also present the ratio between simulated and predicted number of single-single collisions in Table~\ref{tab:list_of_sims} and Fig.~\ref{fig:ratio_data_model}.
We find an excellent agreement between the predicted and simulated number of collisions.

Despite the emergence of more massive and larger stars from collisions, the single-single encounter rate is still able to account for mergers involving unequal mass stars. When compared to the number of single-single collisions only, the analytic model that we use underestimates the simulations results by a factor $1.04$, with a three sigma uncertainty of $\pm0.05$

%\subsection{Collisions occurring post-binary formation}
\subsection{Binary-mediated collisions}
\label{sec:binary_collisions}
We note that in our clusters the density is sufficiently high as to activate the formation of binary stars via close interactions between three isolated single stars. A pair of stars then remain bound in a binary while the third star is ejected, typically taking away more positive kinetic energy than it came in with. 
This sets the scene for subsequent single-binary interactions to occur, in which three stars undergo a chaotic gravitationally-bound interaction within a small volume with a correspondingly high probability of a collision occurring  \citep[see][for more details on the expected probabilities]{leigh12, leigh15, leigh16, leigh18}.
If the collision occurs during an interaction with a third bound star, we classify it as a direct binary collision (see Sec.~\ref{sec:correction_factors}).
Isolated binaries can also be driven to merger due to perturbative encounters with bound or unbound single stars. We present in Fig.~\ref{fig:binary_perturbation} one example for each of these two cases (see Sec.~\ref{sec:t1_2}).

%, correspond to 60 percent of the total number of binary collisions in our simulations.
%We plot in the upper panels the eccentricity of the orbit whereas in the lower panels we plot the semi-major axis and the distance to the closest star. We find that close single-binary encounters produce perturbations to the orbit of the binary. These perturbations drive the eccentricity close to unity, inducing a merger at pericentre. 

In Fig.~\ref{fig:encounter_rates_example_calculation}, we revisit the example simulation considered in the previous section. In the middle panel, the solid and dashed red lines show, respectively, the simulated and predicted numbers of binary mediated collisions, as calculated with Eq.~(\ref{eq:N_1+2}).  
We find that the agreement is better when the number of stars is low $N\leq1000$, except for $N=100$ (see columns 10 and 11 in Table~\ref{tab:list_of_sims}). For our $N=100$ simulations nearly all of the collisions are binary mediated, but there are very few of them. This low number statistics directly impacts the determination of the correction factors $f_{\rm dir}$, $f_{\rm pert}$ and $n_{\rm pert}$ from our higher cadence runs (see Sec.~\ref{sec:correction_factors}), such that they become unreliable for these models (see Table~\ref{tab:correction_factors}).

\begin{figure}
	\includegraphics[width=\columnwidth]{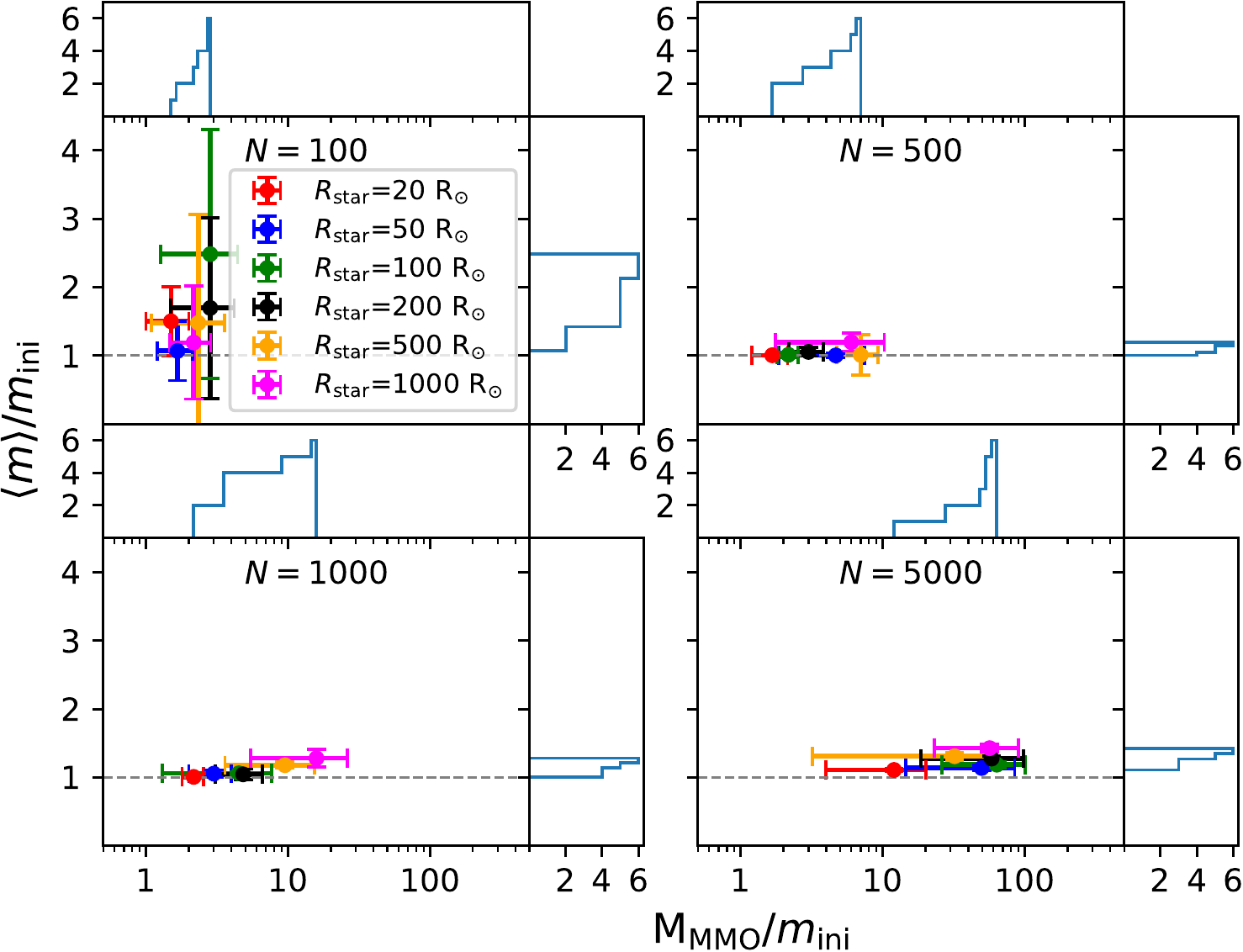}
    \caption{Average stellar mass inside the core as function of the most massive object mass. For low $N$ and small $R_{\rm star}$ there is not a very massive object in the cluster. For larger $N$ and $R_{\rm star}$ the core always contains an object that is much more massive than the rest at the end of the considered time-span. The cumulative histograms at the top and to the right of each panel help to see the widths in the distribution of the mean stellar mass in the core and the MMO mass distribution, respectively.}
    \label{fig:mmo_mmean}
\end{figure}

The better agreement for $N=500$ and $N=1000$ can be explained by the almost unchanged stellar mass distribution. For larger particle numbers (i.e., $N=5000$) the amount of binary collisions increases dramatically towards the end of the simulated time-span, as can be seen in Fig.~\ref{fig:encounter_rates_example_calculation}. We attribute this increased rate of binary mediated collisions to the presence of a very massive object in the cluster core. In Fig.~\ref{fig:mmo_mmean} we plot the average stellar mass inside the cluster core ${\langle m \rangle}$ as function of the mass of the most massive object (MMO). We exclude the MMO from the calculation of ${\langle m \rangle}$. We
show that for $N=5000$ there is always a single object in the core that is much more massive than the rest. 

\begin{figure}
	\includegraphics[width=\columnwidth]{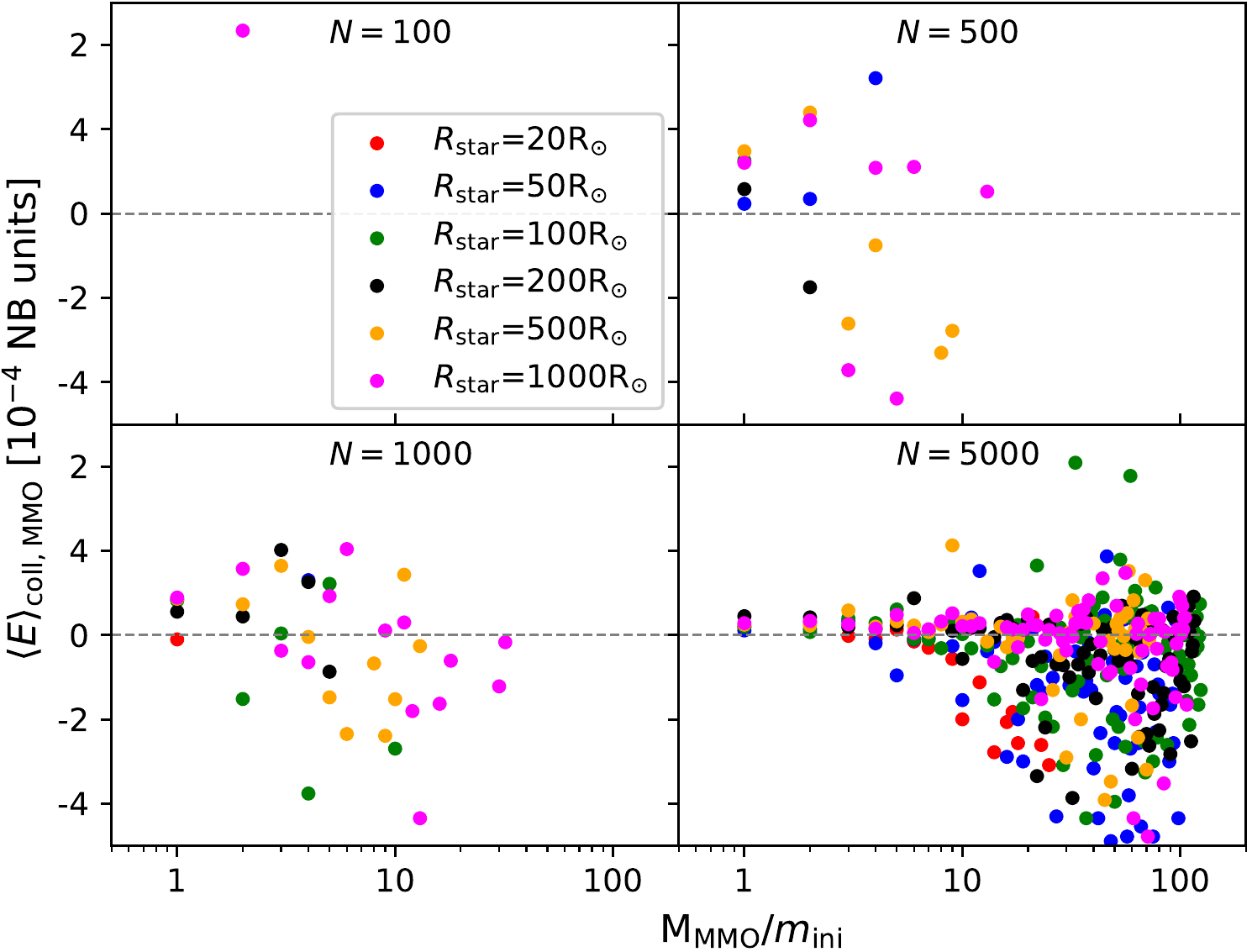}
    \caption{Average kinetic plus potential energy of stars colliding with the most massive object $\langle E \rangle_{\rm coll, MMO}$, as function of the most massive object mass, for different $N$ and $R_{\rm star}$. Hyperbolic collisions are above the dashed grey line whereas binary collisions are below. For larger $R_{\rm star}$ hyperbolic collisions are preferred. Binary collisions are favoured for larger M$_{\rm MMO}/m_{\rm ini}$, which is specially important at late times for $N=5000$. For $N=100$ the rest of the points are around $-100$ NB units, and thus are not shown here.}
    \label{fig:E_coll_MMO}
\end{figure}

In order to illustrate the tendency of the MMO to grow through binary instead of hyperbolic collisions (i.e., single stars that become bound to the MMO on quasi-Keplerian orbits, forming binaries with it), we calculate the average energy of the stars that collide with the MMO as function of its mass. This is done for different values of $N$ and $R_{\rm star}$ and shown in Fig.~\ref{fig:E_coll_MMO}. In this figure, hyperbolic collisions are above the dashed grey line and binary collisions below. The energy is presented in $N$-body (NB) units \citep{Heggie86}. This figure shows that for $N=5000$ and $R_{\rm star} \leq 200~\rm R_{\odot}$ the most massive object in the core preferentially grows via binary-mediated collisions. This helps to explain the discrepancy seen in Fig.~\ref{fig:encounter_rates_example_calculation} and in columns 10 and 11 in Table~\ref{tab:list_of_sims}.

Summarising our results so far, the presence of a very massive object in the core causes the assumptions and approximations going into our analytic model to break down, such that the model should be re-visited and re-constructed to account for these alternative collision channels.  For example, as the MMO grows in mass, we expect secular analytic theories to agree better with the results of the simulations, as perturbations become less important in the deep potential of a central very massive object.

%We plot in Fig.~\ref{fig:fbin_cols} the fraction of binary-mediated collisions that involve the most massive object formed in the cluster. The most massive object is the star with the largest mass at the end of the run. This figure illustrates the fact that for larger particle numbers, most of the discrepancy between or simulated and predicted binary collisions is caused by a very massive object living in the cluster centre, which preferentially forms a binary system with another star before the collision \citep{leigh14}. %Binary collisions with a massive object in the cluster centre are more important for larger $N$ as indicated in Fig.~\ref{fig:fbin_cols}. 
%This is because a massive object is present in the cluster core as a result of previous stellar collisions, and we find that an important fraction of the total number of binary collisions involve the most massive object. This is illustrated in Fig.~\ref{fig:fbin_cols}. This effect becomes more important as time goes on.

%\begin{figure}
%	\includegraphics[width=\columnwidth]{Ratios_AB_large.png}
%    \caption{Ratio of simulated to predicted number of $A+B$ collisions. The predicted number of collisions are obtained with the analytic model described in Sec.~\ref{sec:cols_from_analytic_model}, specifically Eq.~(\ref{eq:NA+B}), and are presented in column 13 in Table~\ref{tab:list_of_sims}. The simulated number of collisions are obtained from our simulations by selecting only collision products whose final mass is 3$m_{\rm ini}$,i.e., selecting the collision product of $A+B$ stars. This information is presented in column 12 in Table~\ref{tab:list_of_sims}.}
%    \label{fig:ratios_AB}
%\end{figure}

%\begin{figure}
%	\includegraphics[width=\columnwidth]{fhyp_N_R.png}
%    \caption{Fraction of hyperbolic collisions that do not involve merger products, as function of $N$ and $R_{\rm star}$ in our simulations. For all our runs, collisions that involve merger products are in general $\lesssim 20\%$ of all the hyperbolic collisions.}
%    \label{fig:fhyp_cols}
%\end{figure}

\begin{table*}
	\centering
	\caption{Numbers and types of collisions for each model. We perform 6 simulations for each model and calculate the mean number of collisions within $t_{\rm sim}$, both from our simulations (column 5) and from our analytic model (column 6). The ratio between the simulated and analytic number of collisions is presented in column 7 and plotted in Fig.~\ref{fig:ratio_data_model}. The number of hyperbolic and binary collisions from the simulations are presented in columns 8 and 10. We also present the number of 1+1 and 1+2 collisions derived from our analytic model in columns 9 and 11.}
	\label{tab:list_of_sims}
	\begin{tabular}{lrrrrrrrrrr} % four columns, alignment for each
		\hline
		Model & $N$ & $R_{\rm star}$ & $t_{\rm sim}$ & $N_{\rm col,sim}$ & $N_{\rm col,analytic}$ & $\frac{N_{\rm col,sim}}{N_{\rm col,analytic}}$ & $N_{\rm Hyp}$ & $N_{1+1}$ & $N_{\rm Bin}$ & $N_{1+2}$ \\ %& $N_{\rm A+B, sim}$ & $N_{\rm A+B, analytic}$\\
		&  & (R$_\odot$) & (Myr) &  &  &  &  &  &  \\	
		\hline
1 & 100 & 20 & 0.400 & 0.333 $\pm$ 0.236 & 0.077 & 4.308 & 0.000 & 0.026 & 0.333 & 0.052 \\ 
2 & 100 & 50 & 0.400 & 0.333 $\pm$ 0.236 & 0.094 & 3.559 & 0.000 & 0.094 & 0.333 & 0.000 \\ 
3 & 100 & 100 & 0.400 & 0.500 $\pm$ 0.289 & 0.145 & 3.439 & 0.333 & 0.145 & 0.167 & 0.000 \\ 
4 & 100 & 200 & 0.300 & 0.667 $\pm$ 0.333 & 0.216 & 3.085 & 0.333 & 0.202 & 0.333 & 0.014 \\ 
5 & 100 & 500 & 0.200 & 0.667 $\pm$ 0.333 & 0.855 & 0.780 & 0.167 & 0.351 & 0.500 & 0.503 \\ 
6 & 100 & 1000 & 0.150 & 0.833 $\pm$ 0.373 & 0.579 & 1.439 & 0.167 & 0.544 & 0.667 & 0.035 \\ 
\\
7 & 500 & 20 & 0.500 & 0.667 $\pm$ 0.333 & 1.444 & 0.462 & 0.167 & 0.254 & 0.500 & 1.189 \\ 
8 & 500 & 50 & 0.900 & 4.333 $\pm$ 0.850 & 3.292 & 1.316 & 1.833 & 1.323 & 2.500 & 1.968 \\ 
9 & 500 & 100 & 0.600 & 2.000 $\pm$ 0.577 & 3.027 & 0.661 & 1.500 & 1.581 & 0.500 & 1.446 \\ 
10 & 500 & 200 & 0.530 & 4.333 $\pm$ 0.850 & 6.302 & 0.688 & 3.000 & 2.637 & 1.333 & 3.666 \\ 
11 & 500 & 500 & 0.400 & 9.333 $\pm$ 1.247 & 9.205 & 1.014 & 6.167 & 4.909 & 3.167 & 4.295 \\ 
12 & 500 & 1000 & 0.370 & 11.333 $\pm$ 1.374 & 14.469 & 0.783 & 9.500 & 8.820 & 1.833 & 5.649 \\ 
\\
13 & 1000 & 20 & 1.349 & 2.667 $\pm$ 0.667 & 2.595 & 1.028 & 2.333 & 1.896 & 0.333 & 0.699 \\ 
14 & 1000 & 50 & 1.349 & 7.833 $\pm$ 1.143 & 8.741 & 0.896 & 5.667 & 4.780 & 2.167 & 3.960 \\ 
15 & 1000 & 100 & 1.000 & 7.000 $\pm$ 1.080 & 8.860 & 0.790 & 4.167 & 5.960 & 2.833 & 2.901 \\ 
16 & 1000 & 200 & 0.900 & 12.000 $\pm$ 1.414 & 14.812 & 0.810 & 9.833 & 10.545 & 2.167 & 4.268 \\ 
17 & 1000 & 500 & 0.700 & 22.667 $\pm$ 1.944 & 25.634 & 0.884 & 17.833 & 17.521 & 4.833 & 8.114 \\ 
18 & 1000 & 1000 & 0.500 & 28.667 $\pm$ 2.186 & 45.490 & 0.630 & 22.333 & 27.668 & 6.333 & 17.822 \\ 
\\
19 & 5000 & 20 & 5.316 & 66.667 $\pm$ 3.333 & 52.655 & 1.266 & 55.167 & 45.064 & 11.500 & 7.591 \\ 
20 & 5000 & 50 & 3.500 & 75.000 $\pm$ 3.536 & 81.020 & 0.926 & 65.333 & 60.352 & 9.667 & 20.668 \\ 
21 & 5000 & 100 & 3.200 & 147.500 $\pm$ 4.958 & 113.954 & 1.294 & 120.500 & 105.549 & 27.000 & 8.405 \\ 
22 & 5000 & 200 & 2.200 & 171.000 $\pm$ 5.339 & 153.362 & 1.115 & 148.333 & 137.331 & 22.667 & 16.031 \\ 
23 & 5000 & 500 & 1.200 & 164.333 $\pm$ 5.233 & 194.490 & 0.845 & 152.667 & 178.430 & 11.667 & 16.060 \\ 
24 & 5000 & 1000 & 0.700 & 205.667 $\pm$ 5.855 & 255.998 & 0.803 & 190.167 & 235.863 & 15.500 & 20.136 \\ 
		\hline
	\end{tabular}
\end{table*}

\begin{table}
	\centering
	\caption{Correction factors for direct and perturbed binary collisions.}
	\label{tab:correction_factors}
	\begin{tabular}{l r r r r r } % four columns, alignment for each
    		\hline
		Model & $N$ & $R_{\rm star}$ & $f_{\rm dir}$ & $f_{\rm pert}$  & $n_{\rm pert}$  \\
		%\hline
		 &  & (R$_\odot$) &  & &  \\
		\hline
1 & 100 & 20 & 0.006 & 0.000 & - \\ 
2 & 100 & 50 & 0.000 & 0.000 & - \\ 
3 & 100 & 100 & 0.000 & 0.000 & - \\ 
4 & 100 & 200 & 0.000 & 0.007 & 3.000 \\ 
5 & 100 & 500 & 0.122 & 0.000 & - \\ 
6 & 100 & 1000 & 0.000 & 0.037 & 4.000 \\
\\
7 & 500 & 20 & 0.017 & 0.000 & - \\ 
8 & 500 & 50 & 0.014 & 0.006 & 2.250 \\ 
9 & 500 & 100 & 0.017 & 0.011 & 11.500 \\ 
10 & 500 & 200 & 0.042 & 0.010 & 1.000 \\ 
11 & 500 & 500 & 0.040 & 0.007 & 1.000 \\ 
12 & 500 & 1000 & 0.071 & 0.026 & 3.75 \\
\\
13 & 1000 & 20 & 0.002 & 0.002 & 1.000 \\ 
14 & 1000 & 50 & 0.016 & 0.002 & 2.000 \\ 
15 & 1000 & 100 & 0.015 & 0.000 & - \\ 
16 & 1000 & 200 & 0.017 & 0.003 & 1.000 \\ 
17 & 1000 & 500 & 0.044 & 0.000 & - \\ 
18 & 1000 & 1000 & 0.108 & 0.010 & 2.667 \\
\\
19 & 5000 & 20 & 0.006 & 0.001 & 1.500 \\ 
20 & 5000 & 50 & 0.015 & 0.009 & 14.679 \\ 
21 & 5000 & 100 & 0.008 & 0.008 & 14.382 \\ 
22 & 5000 & 200 & 0.018 & 0.001 & 6.000 \\ 
23 & 5000 & 500 & 0.020 & 0.010 & 16.578 \\ 
24 & 5000 & 1000 & 0.028 & 0.005 & 6.667 \\ 
		\hline
	\end{tabular}
\end{table}

\section{Applications to observed data and the formation of stellar exotica}
\label{sec:applications}
In this section, we quantify what our results are telling us about the dominant merger/collision channels as a function of environment.  We then go on to confront our results with observed data from Milky Way globular clusters (GCs), to identify the dominant collision mechanism operating in different star cluster environments.  With this information in hand, we move on to making predictions for the properties of stellar exotica that are thought to be the products or progenitors of collisions/mergers, specifically blue stragglers stars and other potential merger products found in galactic nuclei (e.g., the S-stars).

\subsection{Collision timescales and real globular clusters}
\label{sec:Timescales_and_GCs}

We now proceed to compare three different collision rates in the cores of globular clusters (GCs), namely the rate for single-single collisions, the rate for direct single-binary collisions and the rate for binary formation from encounters of three single stars. For the calculation of these rates we assume that the average mass and average stellar radius correspond to Sun-like stars, and for the cluster core radius we take the mean core radius from the GC sample.

\begin{figure*}
	\includegraphics[width=\linewidth]{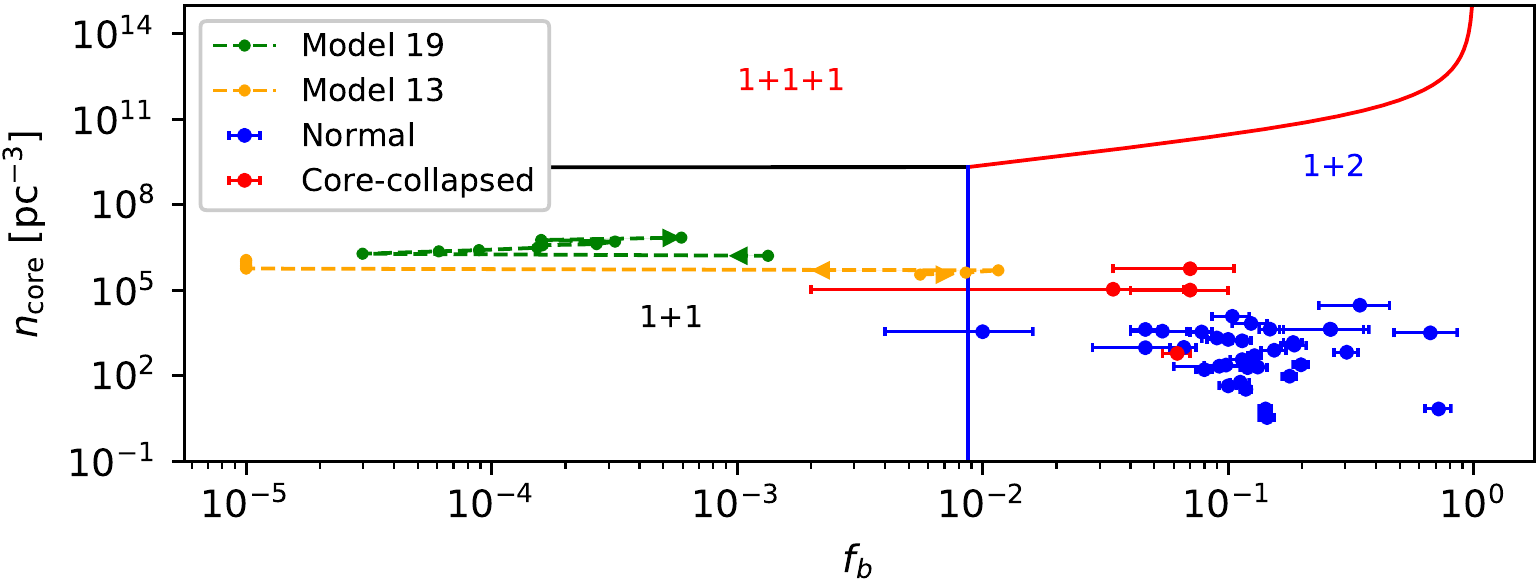}
    \caption{Encounter rates, observed data for globular clusters and data from two of our simulations.
    The solid lines separates the regions where each encounter rate dominates over the others, and the regions are labelled accordingly. We overplot real data for Milky Way Globular Clusters distinguishing core-collapsed clusters (red points) from the rest (blue points). We also overplot data from one of our simulations for Model 19 and Model 13.
    The timescales are calculated assuming a cluster core radius of 2.6~pc (corresponding to the mean of our GC sample) and a mean stellar radius of 1~R$_{\odot}$.}
    \label{fig:Timescales_GC}
\end{figure*}

All these rates depend on the number density of stars and on the binary fraction, and we present in Fig.~\ref{fig:Timescales_GC} the regions of this space where each of these rates dominates over the others.
The timescales for 1+1 and 1+2 interactions are obtained from Eq.~(\ref{eq:tau_1+1}) and (\ref{eq:tau_1+2}), whereas the rate for 1+1+1 interactions is obtained from equation (7.11) in \cite{BinneyTremaine2008} including gravitational focusing.
The red line in Fig.~\ref{fig:Timescales_GC} marks the limit in which the 1+1+1 and 1+2 rates are equal. The black line marks the limit in which the 1+1+1 and 1+1 rates are equal. The blue line marks the limit in which the 1+1 and 1+2 rates are equal.
For computing the 1+2 rate we use the semi-major axis corresponding to the hard-soft boundary. We assume a velocity dispersion of 5~km~s$^{-1}$.

The GC sample we take from \cite{Milone2012}, which contains information about the binary fractions in the cores of the clusters, and we cross-correlate these clusters to the ones in the catalog of \cite{Harris1996}\footnote{An updated version is maintained under: https://cdsarc.unistra.fr/viz-bin/cat/VII/195}.  The latter catalog contains information about the cluster core radius and central luminosity per cubic parsec, which we convert to a core number density $n_{\rm core}$ assuming a mass-luminosity ratio of 1~M$_{\odot}$~L$_\odot ^{-1}$ and mean stellar mass of 1~M$_\odot$.
%are good approximations for the stellar population in the core.

As can be seen in Fig.~\ref{fig:Timescales_GC} the majority of the clusters in this sample fall in the region where binary collisions dominate over hyperbolic collisions.  Hence, once binaries begin to form, they become a significant contribution to the overall rate of collisions and mergers.  The presence of even a few binaries causes the rates for the binary collision channels to dominate over the hyperbolic collision rates.  
We also include in this figure the time evolution of two of our simulations (shown by +the green and orange trajectories). For one of our Model 19 simulations, the cluster always remains in the region dominated by single-single collisions. 

For one of our simulations of Model 13, the cluster starts in the region dominated by single-single collisions.  Some
binaries are formed and $f_b$ increases, thus triggering
binary collisions to dominate for a brief time.  These are, however, a factor $\sim7$ less numerous than hyperbolic collisions. Once a binary forms, it merges or evaporates so quickly that the single-single rate quickly goes back to dominating over the single-binary rate. 
%but in this case a high enough fraction of binary systems is formed and binary collisions dominate during a short period. Then the binary fraction decreases due to collisions/mergers and the cluster core is dominated again by single-single collisions.} 

Overall the rate of hyperbolic collisions is
higher than the rate for single-binary collisions. Over time, however, the altered mass function helps to promote the formation of binary systems between the surrounding stars and the central most massive object, such that binary collisions tend to involve a massive collision product residing in the centre of the cluster.  These binaries tend to merge quickly, due to a perturbation-driven random walk in eccentricity, as shown in the right panel of Fig.~\ref{fig:binary_perturbation}.
%\textcolor{red}{NL: I made edits to the above paragraph, please see if you agree.  Can you change it back to how it was?  I can't compile the pdf or see it on my end, sorry for that.}

\subsection{Blue stragglers}

Blue straggler (BS) stars appear in the cluster colour-magnitude diagram brighter and bluer than the main-sequence (MS) turn-off, where isolated stellar evolution predicts that no stars should be present.  Two main competing mechanisms have been proposed for their formation, namely direct collisions between MS stars and mass transfer onto a MS star in a binary star system.  If most BSs are formed from single-single collisions, then this would predict a correlation between BS numbers and the collision rate, as has been observed for low-mass x-ray binaries \citep{pooley06}.  However, 
there is no clear correlation between BS numbers or relative frequencies, and the collision rate in the cores of globular clusters \citep{Leigh2007}. This, combined with a correlation between cluster core mass and BS numbers, is indirect evidence supporting a binary evolution origin for BSs \citep{Knigge2009,Leigh2011,Leigh2013}. Our results suggest that the complexity in the process of collisions/mergers could play an important role when looking for correlations with collision rates. %, due to the break down of the validity of the ``n$\sigma$v'' approximation.

The process of perturbative single-binary encounters that we see in our simulations could be an important formation mechanism for isolated BSs (i.e., without binary companions) in GC cores. This motivates the development of analytic models like in \cite{Leigh2011}, but more complex as to realistically model the complicated collisional and/or merger processes identified here in high-density clusters with low binary fractions, as illustrated in this paper and \cite{Barrera20}.  Taken at face value, our results predict a higher fraction of isolated BSs relative to BSs with binary companions in clusters with high central densities and low binary fractions. This is in contrast to both the binary evolution channel for BS formation and the mechanism proposed in \cite{Perets2009} for BSs forming from stable hierarchical triples, as they predict BSs with binary companions.  

Let the number of isolated BSs be denoted $N_{\rm BS, iso}$ and the number of BSs with binary companions be denoted $N_{\rm BS,bin}$.  Our results could predict a correlation between the ratio $N_{\rm BS,iso}/N_{\rm BS,bin}$ and both central density and core binary fraction, with this ratio increasing in denser clusters with lower binary fractions.  In principle, this prediction is observationally testable.  Recently, tentative evidence for double BS sequences have been identified in the colour-magnitude diagrams of Galactic globular clusters \citep{Ferraro2009,Dalessandro2013,Simunovic2014}.  The authors propose that one sequence is due to collisions/mergers which would produce isolated BSs, whereas the other sequence is due to binary evolution which would produce BSs with binary companions.  If these double BS sequences are confirmed to be statistically significant, then our results naively predict that the ratio of these two populations should also correlate with central density and core binary fraction. The sample size for observing double BS sequences in globular clusters is currently of order unity.  Hence, a preliminary test of our hypothesis can be done now, but more data would be needed to properly answer the question with statistical significance.  Using the WIYN data from the WOCS collaboration (see \citet{mathieu13}, for example, for more details) would be ideal for this experiment, since it focuses on old open clusters and, at least for clusters like NGC 188, M67 and a few others, extensive studies have been done over the last several decades to thoroughly quantify the issue of cluster membership, and even provide BS binary fractions in some clusters \citep[e.g.][]{mathieu09,geller15a,rain20}.  In order to increase the sample size for the number of clusters with known BS binary fractions, these WIYN data can be combined with the GC data coming from double BS sequences.  Both individual samples continue to grow slowly over time.  The rate could be accelerated even further by looking for double BS sequences in the WIYN data and using speckle interferometry to search for binary companions to individual BSs in open clusters.  Ultimately, however, the data is coming from studies of both open and globular clusters, and it will not be long before we have a reasonable sample size of BS binary fractions to test our prediction that more single BSs should be present in higher-density star clusters due to perturbative mergers of binaries.
Performing a more detailed comparison by confronting our theoretical predictions with real astronomical data will be the focus of future work.

%If we can identify double BS sequences with the WOCS data, then we have a decent sample size and could perform the experiment to obtain a statistically significant result.  
%}}
%Specifically, we expect the ratio $N_{BS,iso}/N_{BS,bin}$ to increase in denser clusters with lower binary fractions (e.g., clusters that are currently in a state of core-collapse or that recently bounced out of one).  
%as isolated binaries are still able to merge after several single-binary perturbations, leaving an isolated BS as the product.

\subsection{The galactic centre and the origins of the S-stars} \label{Sstars}
%\textcolor{red}{BR7: Finally I wrote something here. Please check that sentences make sense and correct grammar mistakes. Feel free to add/remove content}
The innermost parsec in the galactic centre contains a stellar disk
extending out to 0.5~pc \citep{Stostad2015}. This structure is formed by massive young stars with an estimated age of 4-6~Myr and a notably top heavy mass function \citep{Lu13,Bartko2010,Mauerhan2010,Yelda14}. The presence of binary systems in the disk whose components merge due to Lidov-Kozai (LK) oscillations could potentially provide an explanation to observed stellar properties and even the peculiar G2 object \citep{Prodan2015,Stephan2016}. 
Following the same approach adopted by \cite{Prodan2015}, we consider a binary system which in turn is orbiting a central supermassive black hole (SMBH) and we make a crude estimate of the timescale for LK oscillations to operate by using their equation~(2):
%to estimate the timescale for LK oscillations as

\begin{eqnarray}
 \label{eq:LK}
    T_{\rm Kozai} \approx& 2.5\times 10^6 \left( \frac{a_{\rm out}}{0.5 \rm pc}\right)^3 \left( \frac{1 \rm AU}{a_b}\right)^3 \left( \frac{M_b}{2 \rm M_\odot}\right)^{1/2}  \nonumber \\
      &\times \left(\frac{4 \times 10^6 \rm M_\odot}{M_\bullet} \right) \left( \frac{a_b}{1 \rm AU}\right)^{3/2} (1-e_{\rm out}^2)^{3/2} \ \rm yr,
\end{eqnarray}
where $a_{\rm out}$ is the semi-major axis of the binary-SMBH system and $e_{\rm out}$ the eccentricity of the orbit. $a_b$ is the semi-major axis of the binary that orbits the SMBH, $M_b$ is the mass of the binary and $M_\bullet$ is the mass of the SMBH. 

In the galactic centre the stellar density follows a power-law given by $\rho \propto r^{-\gamma}$, and the
gravitational potential is dominated by the central SMBH, thus $v_{\rm rms} = \sqrt{G M_{\bullet}/a_{\rm out}}$. With this, the single binary collision timescale from Eq.~(\ref{eq:tau_1+2}) can be expressed as:

\begin{eqnarray}
\label{eq:t_1_2_GC}
\tau_{1+2, \rm GC} &=& 4.9\times 10^7 (1-f_b)^{-1} f_b^{-1} \left( \frac{a_{\rm out}}{ 0.5 \rm pc} \right)^{1/2 + 2\gamma} \nonumber \\
& & \left( \frac{10^3 {\rm pc^{-3}} \langle m \rangle}{\rho_0} \right)^2 \left(\frac{1 \rm pc}{r_{0}} \right)^3 \left( \frac{4\times 10^6 \rm M_\odot}{M_\bullet}\right)^{1/2} \left( \frac{1\rm AU}{a_b}\right)^2 \nonumber \\
& &\left[ 1+ 53\left( \frac{\langle m\rangle }{0.5 \rm \ M_\odot}\right) \left( \frac{1 \rm \ AU}{a_b}\right) \left( \frac{5\ \rm km\ s^{-1}}{v_{\rm rms,0}}\right)^2 \right]^{-1},
\end{eqnarray}
where, for simplicity, we set $v_{\rm rms,0}=\sqrt{GM_{\bullet}/0.5\rm\ pc}$. We use $\rho_0=5.2\times10^5$~M$_\odot$~pc$^{-3}$, $r_0=0.5$~pc and $\gamma=2$ \citep{Prodan2015}. 

We can now derive the distance from the SMBH at which the timescale for LK oscillations equals the timescale for a perturbed binary collision, for different values of $a_{b}$. We estimate which timescale dominates by doing:
%\textcolor{red}{NL4:  I thikn you could include different curves for different values of e$_{out}$, so we can see how things compete.}

\begin{equation}
    \frac{T_{\rm Kozai}}{0.55} =  \frac{\tau_{\rm 1+2, GC} n_{\rm pert}}{f_{\rm pert} },
\end{equation}
%\textcolor{red}{NL:  Below, I think you need to show explicitly how you calculate the 0.55 factor.  If the window is from 40 to 140, then that is 100/360, so that would be the fraction of binaries oriented in the active LK window.  I think it is 360, since if you flip the binary it becomes retrograde, and with higher-order corrections this can happen.  Maybe ask our collaborators? I guess I am fine sticking with 180 for now, and we'll see what the reviewer says.}
where we introduced the factor 0.55 to account for the fact that the LK mechanism operates for mutual inclinations $\gtrsim 40^{\circ}$, that is, we approximate the fraction of triple systems that are in the active LK window as $100^{\circ}/180^{\circ}$. We set $f_{\rm pert}=0.006$ and $n_{\rm pert}=1.500$ which correspond to our simulation with $N=5000$ and $R_{\rm star} = 20$~R$_\odot$ (see Table~\ref{tab:correction_factors}).

%and the observed value for $M_\bullet$, 

%The timescale for single binary encounters $\tau_{1+2}$ is given by Eq.~(\ref{eq:tau_1+2}) dropping the term in brackets, i.e., neglecting gravitational focusing for simplicity, since we are only interested in an approximation. 

\begin{figure}
	\includegraphics[width=\linewidth]{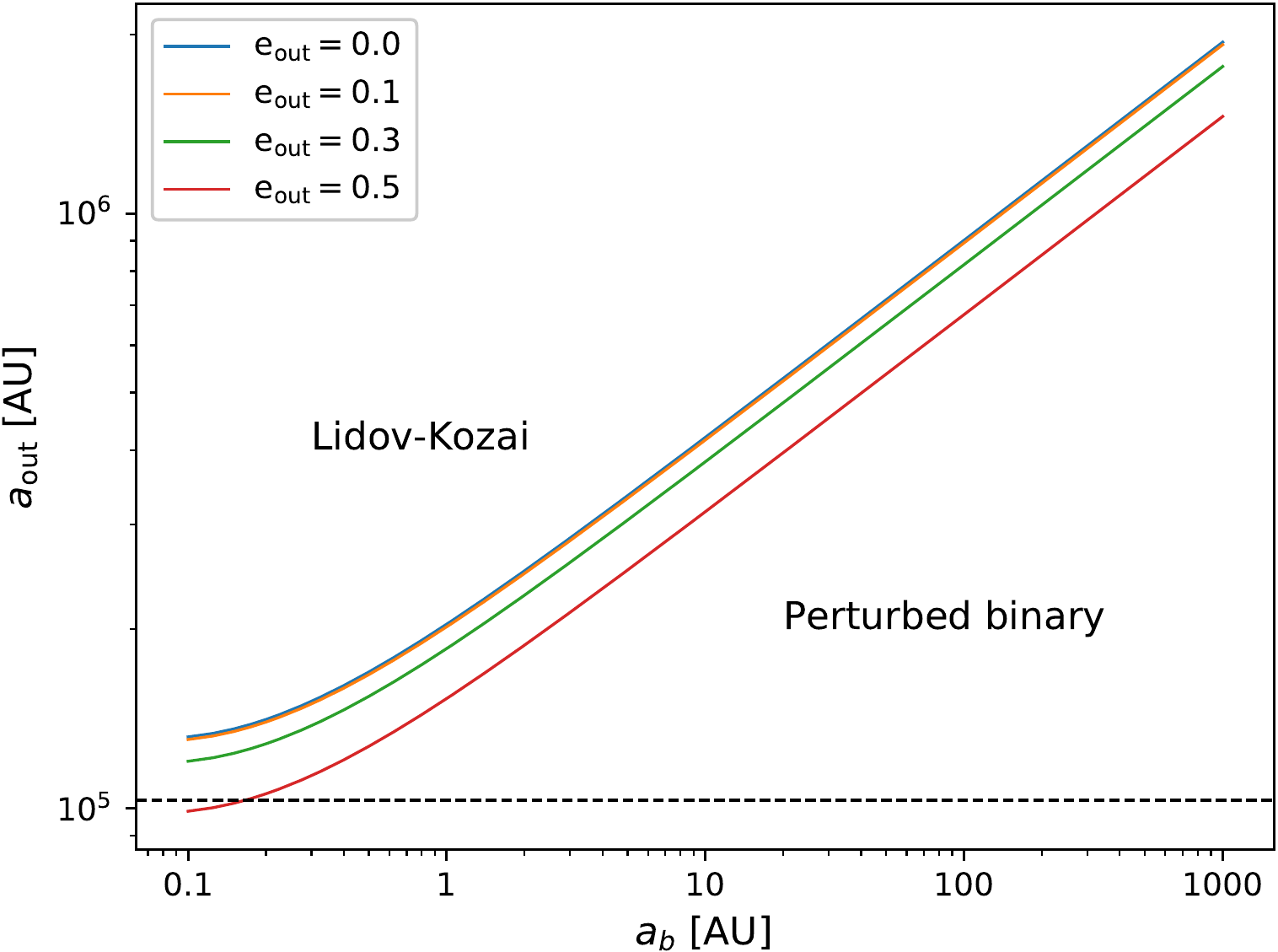}
    \caption{Dominant collisional timescale for binaries in the central galactic disk. $a_{\rm out}$ is the distance from the galactic centre at which the binary orbits, and $a_b$ is the semi-major axis of the binary. The solid couloured lines mark the region where the timescale for one Lidov-Kozai oscillation is equal to the time for perturbed single-binary encounters to produce a merger. Higher eccentricities for the orbit of the binary around the central BH shortens the timescale for perturbed binary collisions compared to collisions from LK oscillations. The regions where each timescale dominates are labelled accordingly. The dashed horizontal line marks the edge of the disk at 0.5~pc.}
    \label{fig:LK_Pert}
\end{figure}

As shown in Fig.~\ref{fig:LK_Pert}, we find that perturbed binary collisions might be more important than LK oscillation-driven collisions when the binary orbits at a distance $\leq 0.5$~pc of the central SMBH, for a wide range of binary semi-major axes. 
The perturbed binary collision scenario that we find in this work thus provides an alternative merger channel that does not require inclined
orbits as in the KL case. This alternative channel requires further investigation to characterize the number of perturbations and the frequency of collisions expected in this environment, which depends sensitively on the assumed density, the mass of the most massive central object and the compactness of the binary. For example, it could occur that only very compact binaries are able to form, and they are relatively insensitive to perturbations.  Indeed, the hard-soft boundary depends on distance from the SMBH, when one is present \citep{leigh16}. 

A more detailed parameter space exploration is required to properly identify how these effects compete in dense environments. Although in Sec.~\ref{sec:binary_collisions} we show that our model is not able to correctly reproduce the number of collisions that involve a central very massive object, we cannot conclude that the model does not work for estimating the number of single-binary interactions for binary systems orbiting such an object. We expect that Eq.~(\ref{eq:t_1_2_GC}) is still able to capture these events, but we aim for a simplified model here that needs to be tested with more sophisticated simulations in the future.
%, however we aim here for a simplified comparison that needs to be confirmed with more sophisticated simulations in the future.}

Here we used the correction factors $f_{\rm pert}$ and $n_{\rm pert}$ for our model that matches better the stellar properties and number density of stars in the galactic centre, however the stellar radii are still too large. We choose not to extrapolate the values of $f_{\rm pert}$ and $n_{\rm pert}$ to smaller radii, and limit ourselves to a very simplified comparison between the two timescales to avoid over-stating the significance of our results (since we have only explored a small subset of the total allowed parameter space of initial conditions).

The take-away message from Fig. \ref{fig:LK_Pert} is that the S-stars could potentially be explained via three-body binary formation forming binaries near to the central SMBH, which then merge due to perturbations from the surrounding high-density $N$-body system.  Hence, Fig.~\ref{fig:LK_Pert} shows that the timescale to drive a merger due to perturbations could be shorter than the timescale for Lidov-Kozai cycles to operate, provided $a_b \lesssim 10$~AU and $a_{\rm out} \lesssim 0.5$ pc.  In this regime, this scenario predicts a top-heavy mass function for stellar populations in the cores of very dense clusters not just due to mergers, but also because it is usually the least massive object that is ejected during three-body binary formation.  The escaping single particle must leave with more kinetic energy than it came in with, such that three-body binary formation would contribute to depleting the central regions of dense environments of lower-mass stars.  Naively, however, this would not explain a disk-like configuration for the orbits of the S-stars, unless star formation first occurred and only formed isolated single stars, which then formed binaries later via 1+1+1 interactions.  To constrain this mechanism, the distribution of orbital eccentricities can be used, since the highly perturbative environment should drive the eccentricity distribution of the S-stars to become supra-thermal.  But detailed $N$-body simulations would first be required to properly quantify the predicted eccentricity distributions for different initial conditions, combinations of BH masses, and so on.

Performing a more rigorous parameter space exploration using sophisticated $N$-body models to properly quantify the post-scattering time evolution of the orbital diffusion in energy- and angular momentum-space for the S-star population will be the focus of future work.  This will be necessary in order to use the observed orbital parameters of the S-star population to test the predictions presented in this section.

\section{Summary and Discussion} \label{sec:discussion}

In this paper, we perform a comparison between analytic calculations for the rate of stellar collisions, to a set of $N$-body simulations that include stellar collisions. Our goal is to test the validity and extent of the mean free path approximation, in a dense dynamically-active star cluster environment.  Our simple numerical models initially consist of equal mass and equal radii particles without stellar evolution, and collisions are treated using the ``sticky star approximation''. This allows us to use simple ``n$\sigma$v'' rates for our analytic model and to avoid complicating effects such as tidal capture, mass loss and tidal disruption.  We focus our analysis to the cluster core, where most simulated collisions occur, and take into account single-single and single-binary collisions.

In general, our analytic model works better for larger $N$ systems where most of the collisions are due to hyperbolic encounters,
but worsens for longer evolution timescales as shown in the middle panel of Fig.~\ref{fig:encounter_rates_example_calculation}.
A longer evolution time produces a larger number of collisions, which in turn promotes the formation of a very massive object in the cluster centre.
As this object grows (preferentially through binary collisions, as we show in Fig.~\ref{fig:E_coll_MMO}) our model begins to break down. This should mark the transition from a chaotic to a more deterministic dynamical evolution, as the contribution from the most massive object to the total gravitational potential increases and the local stellar orbits become increasingly Keplerian. In the limit of large central object masses, the influence radius becomes large (i.e., the distance from the cluster centre of mass to the distance at which point the Keplerian velocity becomes on the order of the local velocity dispersion) and the stellar orbits are typically assumed not to be changing significantly over short time scales, as occurs for low-number chaotic systems.  For our simulations, however, we do not reach this limit, suggesting that perturbative encounters remain important for a large fraction of the parameter space relevant to very dense clusters hosting binary stars (i.e., $n_{\rm core} \gtrsim$ 10$^6$ pc$^{-3}$; see Fig.~\ref{fig:Timescales_GC}). 
In this regime, a loss cone formalism is likely more applicable, but would need to be combined with the standard ``n$\sigma$v'' estimates further out in the cluster where the density is lower and the relevant dynamical timescales, which cause the cluster structure and mass spectrum to change, are longer.

%The reason for the deviations from the mean free path approximation in this regime relates to the stellar motions becoming quasi-Keplerian. This is particularly important when a massive object is present in the core since it tends to grow through binary collisions, as we show in Fig.~\ref{fig:E_coll_MMO}.  Consequently, the assumption of virial equilibrium is also breaking down as the cluster contracts and accumulates merger products formed not only inside the core, but also outside the core which later drift in due to two-body relaxation. 

We include two types of binary collisions in our analytic model, namely direct binary collisions and perturbed binary collisions.  The first type of collisions occur when all three stars become bound, and two or more stars undergo a direct collision during the interaction.  The second type of collision requires that a binary first forms via encounters involving three isolated single stars, but then later merges due to perturbations from the surrounding stellar potential pumping the binary eccentricity to near unity.  For the latter type of collisions, close passages of single stars drive the orbital eccentricity to $e\sim1$ and cause the merger of the binary stars (see Sec.~\ref{sec:t1_2} for details), as we show in Fig.~\ref{fig:binary_perturbation}.

A shortcoming of our model for binary mediated collisions comes from the determination of the correction factors that we introduce to account for direct and perturbed binary collisions. Specifically, for our higher cadence simulations with $N=100$ (see Sec.~\ref{sec:correction_factors}), the number of runs is still so low that we have no binary collisions at all, and hence the correction factors are mostly zero (see Table~\ref{tab:correction_factors}). While this problem is most important for our smallest $N$ simulations, we still have to deal with low number statistics when computing the correction factors for the rest of the models.
This, combined with a low cadence in snapshot outputs, prevents us from accurately exploring the evolution of every binary system and developing a complete understanding of the competition between the effects of the stellar number density, the stellar radius and mass, and the properties of the central MMO on the two binary collision channels that we identify.

We use globular cluster data taken from the literature to determine which of our rates is dominating in which cluster environments.  We obtain core densities from the Harris GC Catalog \citep{Harris1996} and core binary fractions from \cite{Milone2012}.  We plot binary fraction as a function of number density, and segment off those regions of parameter space where each of the 1+1, 1+2 and 1+1+1 rates dominate in Fig.~\ref{fig:Timescales_GC}, as explained in Sec.~\ref{sec:Timescales_and_GCs}.  As is clear, 1+2 interactions dominate in all the globular clusters considered here.  We also overplot the time evolution of two of our simulations in this parameter space.  This indicates that in such dense environments with large radii stars, the single-single collision rate always dominate.  Nevertheless, binaries are still created, but merge or are destroyed relatively quickly without ever reaching a sufficiently high binary fraction for a significant period of time such that single-binary collisions dominate.
%, the relative rates change.  However, we see very clearly that the 1+2 rate typically dominates over the 1+1 and 1+1+1 rates whenever binaries are present, even if the number of binaries is very low.

We suggest that the perturbed binary collisions identified in our simulations could be an alternative merger channel operating efficiently in dense stellar environments where no stable triples can form. This process would produce isolated BSs as opposed to binary stellar evolution, which instead predicts a white dwarf binary companion \citep[e.g.][]{gosnell14,gosnell15}, potentially producing a correlation between the number of isolated BSs and the stellar density.  Specifically, we predict that the ratio of BSs without binary companions to the ratio of BSs with binary companions, or $N_{\rm BS,iso}/N_{\rm BS,bin}$, should increase in denser clusters with lower binary fractions, which offers a possible observational test for this prediction.
By means of a simplified comparison of timescales we show that perturbed binary collisions could be more important than LK induced collisions in the outer parts of the Milky Way central stellar disk. This could have implications for the formation of the S-stars in the Galactic centre, which we quantify qualitatively, however more sophisticated models and simulations are needed to confirm or reject this hypothesis.

Our results have important implications for performing accurate and precise numerical simulations involving collisions and mergers.  Consequently, we caution against blindly using independent analytic approximations in very dense stellar systems. For example, in loss cone theory (see \citet{merritt13} for a detailed review), an analytic model (e.g., a Boltzmann-based diffusion model) is used to compute the torques orbiting objects exert on each other's orbits and hence the timescale for resonant relaxation to operate (see \citet{merritt13b} for a review of resonant relaxation).  The model evolves those orbits by computing the rate of energy and angular momentum exchanged between them, which can then be used to compute the timescale on which stellar orbits diffuse to high eccentricities and pass very close to the system centre of mass, where they would collide with any central massive object.  Our results suggest that this could be an over-simplified analytic model, since it ignores perturbations, which become particularly important when the orbiting bodies are at apocentre.  At high densities, these perturbations become stonger and more frequent.  $N$-body codes are capable of modeling the perturbations and can be used to quantify the competing rates and parameter space.  To the best of our knowledge, however, perturbative effects are still not fully included and any contribution from distant perturbers (e.g., stars in the outskirts of a cluster) are entirely neglected in Monte Carlo simulations for star cluster evolution.  A detailed parameter space study is needed to better understand when the perturbations can be safely ignored, and when they must be included in any analytic model.

\section*{Data Availability}
The data underlying in this work were generated by running {\small NBODY6} in
the computers of the Departamento de Astronom\'ia de la Universidad de Concepci\'on. Data will be shared upon request to the corresponding author.

\section*{Acknowledgements}
We thank the anonymous referee for helpful input.
BR acknowledges support through ANID (CONICYT-PFCHA/Doctorado acuerdo bilateral DAAD/62180013) as well as support from DAAD (funding program number 57451854).  NWCL gratefully acknowledges the generous support of a Fondecyt Iniciaci\'on grant 11180005. DRGS and NWCL acknowledge financial support from Millenium Nucleus NCN19-058 (TITANs). DS thanks for funding via Fondecyt regular (project code 1201280). DS, NCWL and AS acknowledge funding via the BASAL Centro de Excelencia en Astrofisica y Tecnologias Afines (CATA) grant PFB-06/2007.
RSK acknowledges support from the Deutsche Forschungsgemeinschaft (DFG) via the collaborative research centre (SFB 881, Project-ID 138713538) ``The Milky Way System'' (sub-projects A1, B1, B2 and B8) and from the Heidelberg cluster of excellence (EXC 2181 - 390900948) ``STRUCTURES: A unifying approach to emergent phenomena in the physical world, mathematics, and complex data''. He also thanks for funding form the European Research Council in the ERC synergy grant ``ECOGAL -- Understanding our Galactic ecosystem: From the disk of the Milky Way to the formation sites of stars and planets'' (project ID 855130). AS gratefully acknowledges funding support through Fondecyt Regular (project code 1180350)

%%%%%%%%%%%%%%%%%%%%%%%%%%%%%%%%%%%%%%%%%%%%%%%%%%



\bibliographystyle{mnras}
\bibliography{mnras_template} % if your bibtex file is called mnras_template.bib

%\begin{thebibliography}{99}
%\end{thebibliography}

%%%%%%%%%%%%%%%%%%%%%%%%%%%%%%%%%%%%%%%%%%%%%%%%%%

%%%%%%%%%%%%%%%%% APPENDICES %%%%%%%%%%%%%%%%%%%%%

\appendix

\section{Collision timescales with gravitational focusing}
\label{sec:timescales_gravfoc}
For deriving the timescales for single-single collisions ($\tau_{1+1}$) and single-binary collisions ($\tau_{1+2}$) we use the next relations \citep{Leonard1989}:
\begin{eqnarray}
\label{eq:grav_foc}
\sigma_{i + j} &=& \pi p^2 \left[ 1 + \frac{2G(m_i+m_j)}{pv_{\rm rel}^2}\right],\\
\label{eq:gamma}
\Gamma_{i+j} &=& N_i n_j \sigma_{i+j} v_{\rm rel}, \\
\tau_{i+j} &=&\Gamma_{i+j}^{-1}, \\
\label{eq:Nstars}
N_{\rm core} &=&\frac{2}{3}\pi n_0 r_{\rm core}^3, \\
\label{eq:num_dens_core}
n_{\rm core} &=&\frac{n_0}{2}. 
\end{eqnarray}
Here $\sigma_{i+j}$, is the gravitationally-focused cross-section for the interaction between particles $i$ and $j$, $p$ is the pericenter distance for a physical collision, $m_i$ and $m_j$ are the masses of the colliding particles, $N_i$ and $n_j$ are the core number and core number density of particles $i$ and $j$ respectively, $v_{\rm rel}$ is the relative velocity at infinity, $n_0$ is the number density of particles in the cluster centre, $r_{\rm core}$ is the core radius, $n_{\rm core}$ is the mean number density of particles in the core, and $N_{\rm core}$ is the total number of particles in the core.\\

\subsection{Single-single collision timescale}
Consider a physical collision between two single particles with the same mass and radius, $m$ and $R$, respectively. The gravitationally-focused cross-section for such an interaction as calculated with Eq.(\ref{eq:grav_foc}) is:
\begin{eqnarray}
\label{eq:gf_1_1}
\sigma_{1+1} &=& 4\pi R^2 \left[ 1+ \frac{Gm}{Rv_{\rm rms}^2} \right],
\end{eqnarray}
where for a Maxwellian velocity distribution the relative velocity between the two particles is equal to the square root of two multiplied by the root mean square velocity of the particles, i.e, $v_{\rm rel}=\sqrt{2}v_{\rm rms}$.
 The number of single stars in the core is $N_s=N_{\rm core }(1-f_b-f_t)$, where $f_b$ and $f_t$ are the fraction of binary and triple systems in the core, i.e., the number of binary and triple systems in the core, divided by the number of single stars, plus binary systems, plus triple systems in the core. By using Eq.(\ref{eq:Nstars}) and Eq.(\ref{eq:num_dens_core}) we can show that:
  \begin{eqnarray}
 N_sn_s &=&N_{\rm core}^2  (1-f_b-f_t)^{2} \frac{3}{4\pi r_{\rm core}^3}, \nonumber \\
 \label{eq:1st_factor}
 N_sn_s &=&\frac{\pi}{3} \ n_0^2 \ r_{\rm core}^3 \ (1-f_b-f_t)^{2}.
 \end{eqnarray}
 Now, combining Eq.(\ref{eq:gamma}), (\ref{eq:gf_1_1}), (\ref{eq:1st_factor}), and setting $v_{\rm rel} = \sqrt{2}v_{\rm rms}$ we obtain:
 \begin{eqnarray}
 \label{eq:gamma1_1}
 \Gamma_{1+1} &=&\sqrt{2}\frac{4 \pi^2}{3} (1-f_b-f_t)^{2} \ n_0^2 \ r_{\rm core}^3 \ R^2 \times  \nonumber \\ 
 & & \ v_{\rm rms} \ \left[ 1+ \frac{Gm}{Rv_{\rm rms}^2} \right].
 \end{eqnarray}
 Finally, inverting Eq.(\ref{eq:gamma1_1}), inserting some typical values for star clusters, and using $n_{\rm core}$ instead of $n_{\rm 0}$, we get:
 \begin{eqnarray*}
\tau_{1+1} &=& 8.3\times 10^{13} \ (1-f_b-f_t)^{-2} \left( \frac{10^3\ \rm pc^{-3}}{2n_{\rm core}}\right)^2   \times \\ 
 & & \left( \frac{1 \ \rm pc}{r_{\rm core}}\right)^3 \left( \frac{5\ \rm km\ s^{-1}}{v_{\rm rms}}\right) \left( \frac{0.5 \ \rm R_\odot}{R}\right)^2 \times \\
 & & \left[ 1+ 7635\left( \frac{m}{0.5 \rm \ M_\odot}\right) \left( \frac{0.5 \rm \ R_\odot}{R}\right) \left( \frac{5\ \rm km\ s^{-1}}{v_{\rm rms}}\right)^2 \right]^{-1} yr.
\end{eqnarray*}
\\

\subsection{Timescale for single-binary collisions}
Consider now an encounter between a single star and a binary system, with all the stars having the same mass $m$ and radius $R$. The gravitationally-focused cross-section for such an interaction, assuming a pericenter distance equal to the semi major axis of the binary $p=a_b$, as calculated with Eq.(\ref{eq:grav_foc}) is:
\begin{eqnarray}
\label{eq:gf_1_2}
\sigma_{1+2} &=& \pi a_b^2 \left[ 1+ \frac{3Gm}{a_bv_{\rm rms}^2} \right],
\end{eqnarray}
 where for a Maxwellian velocity distribution the relative velocity between the two particles is equal to the square root of two multiplied by the root mean square velocity of the particles, i.e, $v_{\rm rel}=\sqrt{2}v_{\rm rms}$.
 The number of single stars in the core is $N_s=N_{\rm core}(1-f_b-f_t)$ and the number of binary systems in the core is $N_b=N_{\rm core}f_b$, where $f_b$ and $f_t$ are the fraction of binary and triple systems in the core. By using Eq.(\ref{eq:Nstars}) and Eq.(\ref{eq:num_dens_core}) we can show that:
 
  \begin{eqnarray}
 N_sn_b &=&N_{\rm core}^2  (1-f_b-f_t)f_b \frac{3}{4\pi r_{\rm core}^3}, \nonumber \\
 \label{eq:1st_factor_bin}
 N_sn_b &=&\frac{\pi}{3} \ n_0^2 \ r_{\rm core}^3 \ (1-f_b-f_t)f_b.
 \end{eqnarray}
 
 Now, combining Eq.(\ref{eq:gamma}), (\ref{eq:gf_1_2}), (\ref{eq:1st_factor_bin}), and setting $v_{\rm rel} = \sqrt{2}v_{\rm rms}$ we obtain:
 \begin{eqnarray}
 \label{eq:gamma1_2}
 \Gamma_{1+2} &=&\sqrt{2}\frac{\pi^2}{3} (1-f_b-f_t)f_b \ n_0^2 \ r_{\rm core}^3 \ a_b^2 \times \nonumber \\
 & & v_{\rm rms} \ \left[ 1+ \frac{3Gm}{a_bv_{\rm rms}^2} \right].
 \end{eqnarray}
 Finally, inverting Eq.(\ref{eq:gamma1_2}), inserting some typical values for star clusters, and using $n_{\rm core}$ instead of $n_{\rm 0}$, we get:

\begin{eqnarray*}
\tau_{1+2} &=& 1.8\times 10^9 \ (1-f_b-f_t)^{-1} \ f_b^{-1} \left( \frac{10^3\ \rm pc^{-3}}{2n_{\rm core}}\right)^2  \times \\
& & \left( \frac{1 \ \rm pc}{r_{\rm core}}\right)^3 \left( \frac{5\ \rm km\ s^{-1}}{v_{\rm rms}}\right) \left( \frac{1 \ \rm AU}{a_b}\right)^2 \times \\
 & & \left[ 1+ 53\left( \frac{m}{0.5 \rm \ M_\odot}\right) \left( \frac{1 \rm \ AU}{a_b}\right) \left( \frac{5\ \rm km\ s^{-1}}{v_{\rm rms}}\right)^2 \right]^{-1} yr.
\end{eqnarray*}

%\subsection{Timescale for $A+B$ collisions}
%Including gravitational focusing, the encounter rate
%between stars of type $A$ and stars of type $B$ ($m_{A} = 2m_{B}$, $R_A=1.26R_B$), including gravitational focusing, is \citep{leigh12}:
%\begin{eqnarray}
% \label{eq:AplusB}
%    \Gamma_{A+B} &= &{N_A\choose 1} {N_B\choose 1} \frac{12 \sqrt{2} m_B^{3/2} (R_A + R_B)^2 |E|^{7/2}}{G^3 m_A^{1/2} {\langle m \rangle} M^{13/2}} \left( 1+ \frac{6Gm_B}{2.26R_Bv_{\rm rms}^2}\right) \nonumber
%\end{eqnarray}


%%%%%%%%%%%%%%%%%%%%%%%%%%%%%%%%%%%%%%%%%%%%%%%%%%

% Don't change these lines
\bsp	% typesetting comment
\label{lastpage}
\end{document}